\documentclass[compsoc, conference, a4paper, 10pt, times]{IEEEtran}

\usepackage{cite}
\usepackage{amsmath,amssymb,amsfonts}
\usepackage{algorithmic}
\usepackage{graphicx}
\usepackage{textcomp}
\usepackage{bmpsize}
\usepackage{xcolor}
\usepackage{lipsum}
\usepackage[colorlinks=false,urlcolor=black]{hyperref}
\def\BibTeX{{\rm B\kern-.05em{\sc i\kern-.025em b}\kern-.08em
    T\kern-.1667em\lower.7ex\hbox{E}\kern-.125emX}}
\usepackage{tikz}
\usepackage{amsmath}
\usepackage{longtable}
\usepackage{adjustbox}
\usepackage{ltablex}
\usepackage{placeins}
\usepackage{comment}
\usepackage{float}
\usepackage{textgreek}
\usepackage{filecontents}

\usepackage{multirow}
\usepackage{graphics}
\usepackage{fixltx2e}
\usepackage{subfig}

\usepackage{relsize}   
\begin{document}

\title{SoK: Content Moderation in Social Media, from Guidelines to \\ Enforcement, and Research to Practice
}

\author{
\IEEEauthorblockN{Mohit Singhal\IEEEauthorrefmark{1}
Chen Ling\IEEEauthorrefmark{2}
Pujan Paudel\IEEEauthorrefmark{2}
Poojitha Thota\IEEEauthorrefmark{1}
Nihal Kumarswamy\IEEEauthorrefmark{1}}
\IEEEauthorblockN{Gianluca Stringhini\IEEEauthorrefmark{2}
Shirin Nilizadeh\IEEEauthorrefmark{1}}

\IEEEauthorblockA{\IEEEauthorrefmark{1} The University of Texas at Arlington}
\IEEEauthorblockA{\IEEEauthorrefmark{2} Boston University}
}
\maketitle

\begin{abstract} 
Social media platforms have been establishing content moderation guidelines and employing various moderation policies to counter hate speech and misinformation. 
The goal of this paper is to study these community guidelines and moderation practices, as well as the relevant research publications, to identify the research gaps, differences in moderation techniques, and challenges that should be tackled by the social media platforms and the research community. 
To this end, we study and analyze fourteen most popular social media content moderation guidelines and practices, and consolidate them. 
We then introduce three taxonomies drawn from this analysis as well as covering over two hundred interdisciplinary research papers about moderation strategies. 
We identify the differences between the content moderation employed in mainstream and fringe social media platforms. 
Finally, we have in-depth applied discussions on both research and practical challenges and solutions.

\end{abstract}

\section{Introduction}
Social media and online communities allow individuals to freely express opinions, engage in interpersonal communication, and learn about new trends and new stories. 
However, these platforms also create spaces for uncivil behavior and misinformation. 
Uncivil behavior is defined as explicit language, derogatory, or disrespectful content, which has become native on online platforms~\cite{anderson2016toxic,hwang2008does,zannettou2020measuring}. 
Misinformation is defined as false or inaccurate information~\cite{wu2016mining}, which has become rampant on social media platforms. 

Uncivil behaviors like online harassment have a severe impact on users; social media provides anonymity, which can lead to disinhibition, deindividuation, and a lack of accountability, that can lead to anxiety and depression; or even suicide~\cite{suler2004online,brown2018so,fichman2019impacts,ashktorab2016designing}. Misinformation significantly impacts users with undesirable consequences and wreaks havoc on wealth, democracy, health, and national security~\cite{hassan2017toward}. Misinformation, conspiracies, and coordinated misinformation campaigns were prevalent throughout the COVID-19 pandemic~\cite{memon2020characterizing,sharma2020identifying}. 
Such low-quality posts can also drown out useful content and exhaust the limited attention of users~\cite{kiesler2012regulating}. 

Since the early incubation of online communities, scholars and community managers alike reminisce over how to best manage online content and how to enable constructive, civil conversations among the users~\cite{bruckman1994approaches,dibbell1994rape}. 
However, there is no unified method for content moderation among the different social media platforms. 
Some employ more restrictive rules, while others emerge promising no or minimal moderation. 
For example, fringe social media platforms, such as Parler and Gab, have very minimal restrictions and they rarely ban users~\cite{conservfav,flock}. 

Content moderation consists of several levels, including community guidelines, techniques to detect violations, and then policy enforcement. On each platform, content moderation is not constant but evolves as new challenges emerge, or it becomes clear that the in-use methods are not sufficient to protect information integrity. For example, during the 2020 Presidential elections and COVID-19 with the emerge of huge amount of misinformation and fake news on Twitter, the platform started using warning labels on posts to counter such content~\cite{twitter-covid19,twitter-elections}. 
In this paper, we study and categorize the topics covered in content moderation research and investigate the current state of content moderation on several social media platforms, focusing on the enforcement of moderation policies and also the community guidelines and how platforms define and moderate different types of content. With this analysis, we aim to obtain a comprehensive vision of content moderation from the points of view of both the research community and real-world practices.
In particular, we try to answer the following research questions: 
\noindent \textbf{RQ1}: In what aspects and how does the research community study content moderation? 
What are the research gaps that need to be filled? 
\noindent \textbf{RQ2}: How does content moderation work in practice?
What content do different social media platforms try to moderate, and how are the content moderation policies defined, implemented, and enforced? 
What are the practical and research gaps that need to be filled?

Studying and investigating these two research questions helps us in understanding all major components of the content moderation framework. Systemizing publications studying these components, and identifying their overlaps and differences, can help understand the research gaps (RQ1) in each component. Furthermore, no other work has provided a systemization of content moderation which are practiced in social media platforms (RQ2). 
To answer these research questions, we collate more than two hundred plus research papers describing the ever-growing changes to the content moderation practices of social media platforms and their impact on the end-user, and also investigate fourteen social media platforms to understand the current state of content moderation practices. With these analyses, we create three granular taxonomies, which explain and categorize the moderation policies, the types of content that is moderated, and the comprehensiveness of the provided community guidelines. To the best of our knowledge, ours is the first paper that systematizes content moderation policies in social media platforms. Papers studying social media as a business (i.e., platforms as exploitative data-gathering/ surveillance systems for advertising) are out of scope of this paper.  

In our analysis, we found that there are still inconsistencies in the way social media platforms moderate content and how they define the guidelines. For example, while certain categories of abuse are banned across all the platforms, there is no consensus among platforms on what content to moderate and what not to. Hence, moderation happens arbitrarily. From our analysis of previous literature on \emph{misinformation} and \emph{hate speech} detection and the effectiveness of guidelines and enforcement methods, we identified and discussed several research gaps. We believe that our findings will help not only the computer research community but also the social media companies, to  create a more inclusive and transparent moderation process.

\section{Content Moderation}
Moderation is the governance mechanism that structures participation in an online community to facilitate cooperation and prevent abuse~\cite{grimmelmann2015virtues}.
It determines which posts and users are allowed to stay online and which are removed or suspended, how prominently the allowed posts are displayed, and which actions accompany the content removals, i.e. chance to appeal the decision~\cite{costodian}. 
\begin{figure}[t]
    \centering
    \includegraphics[width=0.85\columnwidth]{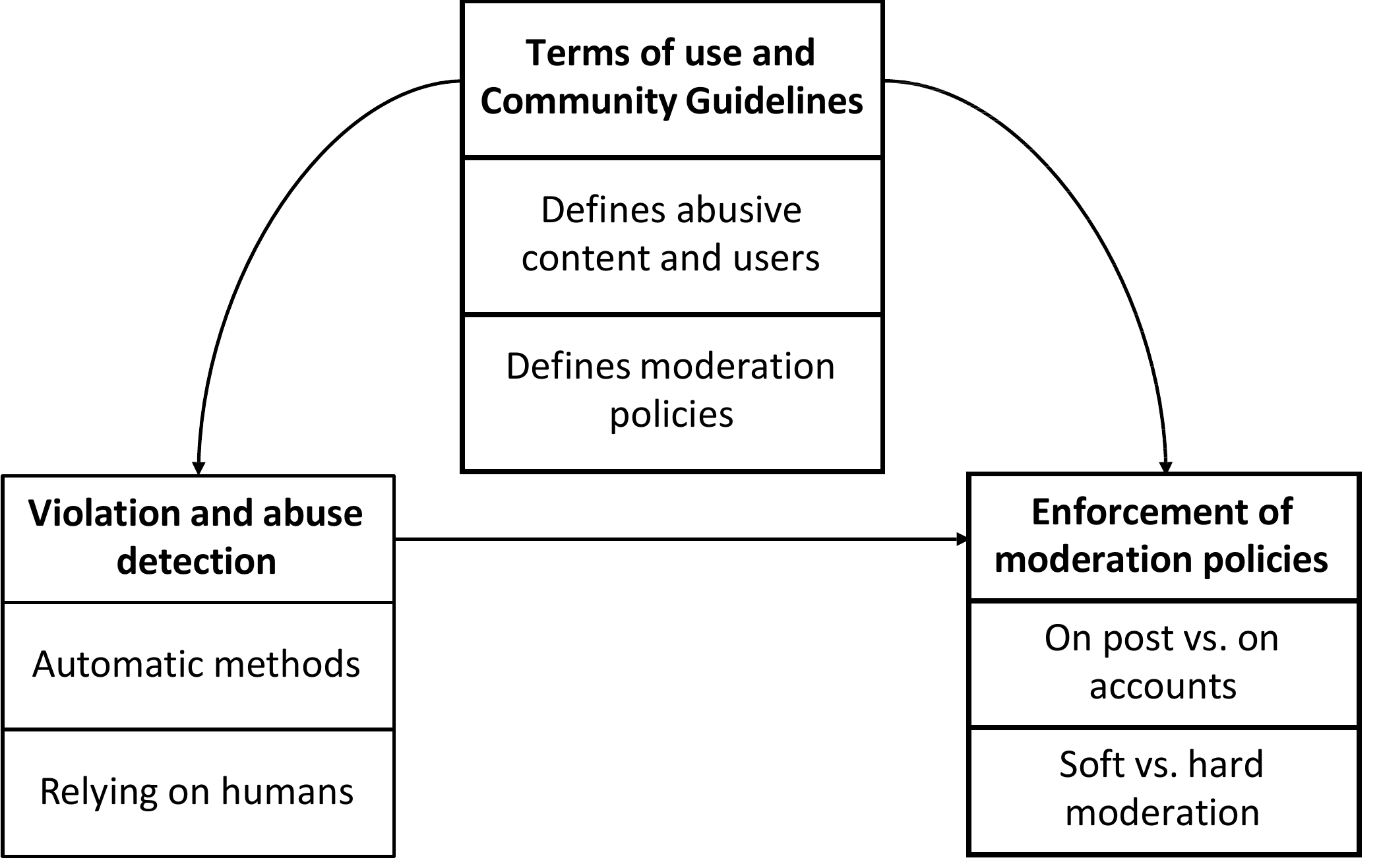}
    \caption{Content Moderation Framework}
    \label{fig:cm-t}
\end{figure}
Figure~\ref{fig:cm-t} shows the content moderation framework, which we generated based on an extensive review of previous works and community guidelines. 
This framework includes three interconnected components: (1)~terms of use and community guidelines which define abusive content and behavior, as well as moderation policies, (2)~violation and abuse detection methods, and (3)~enforcement of the moderation policies. 
Terms of use and community guidelines show how the platforms detect violations and abuse, and  dictate the policies that the platform tries to enforce. 
Throughout this paper, we discuss the content moderation and literature with regards to each of the components. 

\subsection{Terms of Use and Community Guidelines} 
Social media platforms define their content moderation policies alongside their privacy policies, copyright etc., in their \emph{terms of service} and \emph{community guidelines}. 
Typically, whenever a user joins a social media platform, they come across \emph{terms of service} and they are required to provide affirmative agreement to register an account and use the platform. 
Terms of service are not particularly useful as an education tool as they are likely written in \emph{legal} terms and previous works have found that users rarely read them~\cite{steinfeld2016agree}. 
In order for the users to better understand the rules, these platforms provide them as \emph{community guidelines/ standards}, which are established in layman's terms. These terms and guidelines specify the types of content that are prohibited on the platform, and the actions that will be taken once a violation is detected (i.e., moderation policies). 
Some examples of such content are \emph{child sexual exploitation, terrorism, and pornography}, which by law are required to be removed. 
Other types of prohibited content are \emph{malicious content}, which includes spam, malware, and phishing URLs purposefully spread on social media platforms to gain more victims~\cite{zhang2012detecting,cao2015detecting,singhal2021prevalence,singhal2019analysis}. While there are numerous works on the moderation of such content, these works are out of the scope of this paper. 
Instead, we investigate research that focus on content, such as \emph{fake news}, \emph{misinformation}, and \emph{offensive and hate speech}. These are emerging and gray areas, with no universally accepted definitions, which contribute to controversial discussions on the trade-off between moderation and freedom of speech. 

\subsection{Violation and Abuse Detection}
To detect violations as well as abusive content and behavior, social media platforms rely on a combination of human moderators and automated algorithms, which include heuristic-based and rule-based techniques as well as sophisticated machine learning-based models. 

\textbf{Human-based Moderation: } Human moderators can be paid workers or volunteers who not only identify and vet abusive content, but may also enforce the moderation policies, e.g., by directly removing such content. 
Some platforms, like Twitter, Facebook, etc., employ a large group of freelancers, who work on contracts~\cite{con,roberts2016commercial}, while others, for example, Reddit, rely on volunteer moderators, that are typically selected from the most actively involved users in the community
~\cite{matias2016civic,jhaver2019human}. 
Reddit also gives moderators the power to create and enforce local rules that sit below the broader rule set of the platform. 
4chan is similar to Reddit, where there are additional rules specific to each board. 
Some platforms, such as Facebook, Youtube, Snapchat, etc., also rely on users, to flag content that they deem unfit for the readers or they violate the community guidelines~\cite{crawford2016flag}. 
These platforms may then use heuristics or models to employ  moderation policies based on these flags, e.g., they might have a simple rule that a post is deleted if enough users flag it, or they might send such flagged content to their human moderators, who closely vet and enforce the  policies. 

\textbf{Algorithmic-based Moderation: } 
Human-based moderation though effective comes at a cost of time, labor, and liability for the host company, and also at the emotional cost of the workers since they are at the frontlines~\cite{crawford2016flag,roberts2016commercial}. 
To minimize such trade-offs, social media companies are increasingly deploying rule-based techniques and machine learning models to automate the process of identifying problematic material~\cite{mode-2}. 
For detecting different types of violation and abusive content, social media platforms develop a different set of algorithms and methods.   
For example, YouTube relies heavily on its automated flagging systems to remove offensive comments~\cite{youtr}.
Perceptual hashing~\cite{niu2008overview} is used to automatically detect pornographic images and videos, adult nudity, etc., which involves fingerprinting some perpetually salient feature set of the content. 
Combating terrorism content is challenging and requires instantaneous removal from the platform. Companies such as Facebook, Google, Twitter, and Microsoft created a group called Global Internet Forum to Counter Terrorism (GIFCT), which maintains a Shared Industry Hashed Database (SIHD)~\cite{fac-gif,GIFCT}. 
 To detect misinformation in images, Meta has deployed SimSearchNet++, which can detect variations in an image with high precision. Pre-trained NLP language models such as BERT~\cite{mozafari2020hate,alatawi2021detecting,saha2019hatemonitors}, RoBERTa~\cite{liu2019roberta} \& XLM-R~\cite{conneau-etal-2020-unsupervised} have been recently proposed to detect hate speech and are used by Facebook~\cite{fac-misinfor,fac-hates}. 
In academia, most hate speech and toxicity detection studies use Google's Perspective API to detect toxicity in text~\cite{jigsaw2018perspective}. 
The API uses machine learning techniques and a manually curated dataset of text to identify toxicity. 
Most social media platforms do not publicly share the details of their tools used to detect misinformation or offensive content, however, in the literature, AI-based models are proposed to detect specific types of misinformation about COVID-19 and vaccinations~\cite{shahi2021exploratory,kouzy2020coronavirus}, security tools~\cite{singhal2021prevalence}, hate speech, and offensive comments~\cite{zhang2016conversational,davidson2017automated}, fake reviews~\cite{martens2019towards,mukherjee2012spotting}, etc. 

\textbf{Human-in-the-loop moderation:}
Relying solely on algorithms to moderate also has some limitations, e.g., the performance of AI-based models can significantly drop for out-of-distribution examples that differ from the training data~\cite{beede2020human,jia2017adversarial}. Hence, some studies have proposed using human-in-the-loop moderation~\cite{lai2022human,mackeprang2019discovering}, which refers to interactive training paradigm where the AI receives input from the human to improve its performance. 
Moreover, some legislation, like \emph{The European Union General Data Protection Regulation (GDPR)} has mandated human-in-the-loop moderation.
Article 22 \& recital 71 of GDPR guarantees that a decision should not be based solely on automatic systems, if the decision is challenged, it would be subjected to human review~\cite{gdpr,goodman2017european,gdpr-1}.

\subsection{Content Moderation Policy Enforcement}
Terms of service and community guidelines may also define the policies that the platforms enforce when violations of service or abusive behavior are detected. 
These policies can take place at various levels, e.g., at the post level vs. account level~\cite{myers2018censored,merrer2020setting}, and also specify various consequences, e.g., showing a warning vs. removing the abusive account. We found two popular moderation policies that platforms use: \emph{hard} and \emph{soft} moderation. 

\textbf{Hard Moderation: } Hard moderation is the most severe way a platform enforces its policies, which removes the content or entities from the platform~\cite{chandrasekharan2020quarantined,saleem2018aftermath,ribeiro2020does}, when a violation of community guidelines occur.
Then, other users cannot access the content or connect with the abusive account. 
Platforms provide an option for the affected users to appeal for the content that might be wrongly removed by the platform~\cite{fbb-apeals,tww-apeals}.

\textbf{Soft Moderation: } Nowadays, soft moderation is the platforms' first line of defense to counter content that violates their content guidelines. 
In soft moderation, platforms do not remove the content. However, they aim to inform the users about potential issues with the content by adding a warning label, substantiating the post with labels in order to inform and educate the users, or limiting the reach of the doubtful content by putting it in quarantine~\cite{zannettou2021won,mena2020cleaning,geeng2020social}. 
Soft moderation has recently received a lot of attention, both in academic circles and by social media platforms, because of the broader effects and effectiveness of this approach.
For example, since the invasion of Ukraine by Russian forces, Twitter started adding labels to tweets from Russian \& Belarusian state-affiliated media websites~\cite{politico-rus}.
Reddit started \emph{quarantining} r/NoNewNormal, a subreddit that is generally antimask, anti-vax, and is against any governmental COVID-19 restrictions across the world~\cite{subredditq}. 
In quarantine, users are shown a warning page, and they have to make a deliberate choice to view the content. Facebook, Instagram, and Twitter have a \emph{strike system} in place to discourage users from posting misleading false content~\cite{twitter-covid19,facbenook-restriction}. 
For example, if a user gets two or three strikes, then they will not be able to access their account for 12 hours. 
These steps are still considered as soft moderation, as the user is not permanently banned from the platform. 

\section{Content Moderation in Literature} 
\label{lit}
To answer our first research question, we examined the last five years of research from the top security conferences (1.45\%), such as IEEE Security and Privacy, USENIX Security, NDSS, ACM CCS, IEEE Euro S\&P, top data mining conferences (4.5\%), such as KDD, WSDM, CIKM, and ICDM, top machine learning conferences (4\%), such as AAAI AI, IEEE/CVF,
top linguistic conference (13.5\%), such as ACL Anthology, and also from the top HCI and social science conferences and journals (35.7\%), such as CHI, CSCW, ICWSM, The Web Conference, New Media \& Society, Political Behaviour, etc. 
While this distribution shows that content moderation is an interdisciplinary topic, our paper can spark further research from the security community. We list the distribution of papers in Table~\ref{conf-list} in Section~\ref{ekk}.
In our study, we focus on publications investigating topics related to content moderation practices for countering online hate speech, online harassment, trolling, cyberbully or in general online toxicity as well as misinformation, disinformation and fake news. We used the snowballing approach~\cite{goodman1961snowball} for finding relevant papers. We started our search by a set of keywords (i.e., content moderation, hate speech, misinformation, detection, etc.), and expanded our search, by studying their references and related work sections. 
The overall set of keywords are: \emph{content moderation} $+$ (social media, tools, effectiveness, engagement, support, removal, removal comprehension, bias), (soft, hard) moderation, and (hate speech, misinformation) (detection, identification, system, automatic, methods, tools, NLP). 

To create categories for papers about detection of hate speech and misinformation as well as the moderation studies, three authors classified the papers using the open coding process~\cite{glaser2017discovery}. The three authors, independently studied the papers to determine themes, sub-themes, and the  approaches and methods that are employed to address a specific task. 
To find the agreement score, we gave a value of 1, if three authors had a perfect agreement on the categories, otherwise, we divided the number of matched categories by the number of possible categories. Using this methodology, we found substantial agreements of 75\% and 70\% for groupings of detection methods and moderation studies, respectively. 
In Section~\ref{dtt}, we list the identified categories for the publications that are about the detection techniques focusing on \emph{Hate Speech} and \emph{Misinformation}, and in Section~\ref{mpp}, we list the identified categories for publications regarding moderation policies. If a study fits in multiple categories e.g., using content based features and DNNs, then we label that study as hybrid approach.

\subsection{Detection Techniques} 
\label{dtt} 
Some recent surveys~\cite{fortuna2018survey,schmidt2017survey,zhou2020survey,guo2019future,islam2020deep,yin2021towards} have reviewed detection techniques with respect to either hate speech or misinformation. However, these works focused on specific methods (such as NLP, graph-based, etc.). Our work, in contrast, studies methods detecting both types of abuse on social media and, more importantly, studies these methods considering the entire moderation ecosystem. We systemize the detection techniques based on their methods, grouping them by similarities and differences between the underlying methods (e.g., propagation-based techniques, lexicon-based techniques, etc.) under a common taxonomy. We identified five and seven broad categories for hate speech detection and misinformation detection, respectively, which is shown in Table~\ref{metrics-detection} in Section~\ref{app}. 
We found that both hate speech detection and misinformation detection systems equally use features derived from the textual part of the social media posts, such as TF-IDF, Part of Speech (POS) tags, etc.~\cite{davidson2017automated,della2018automatic}. Lexicon-based methods are more common in hate speech detection~\cite{basile2019semeval,vargas2021contextual}. Propagation structure, crowd intelligence-based methods, and knowledge-based methods are popular in misinformation detection due to their spread on social media platforms~\cite{sampson2016leveraging,qian2018neural}. We found that both hate speech and misinformation detection systems have leveraged the advances of Deep Neural Networks (DNNS) advances, eliminating the need for feature engineering and domain expertise~\cite{modha2019ld,wu2018tracing}. Misinformation and hate speech have evolved across modalities, including images and videos. We found methods combine these different modalities to identify misinformation (e.g., fauxtography) and hate speech (e.g., hateful memes)~\cite{pramanick2021momenta,wang2018eann}.

\textbf{Hate Speech Detection} 
Previous works have extensively used Machine Learning (ML) techniques to detect hate speech, offensive language and toxicity online. 
Some works~\cite{frenda2019online,vidgen2020detecting} have proposed detection methods for specific types of hate speech such as misogyny, sexism, Islamophobia, etc. Other works~\cite{davidson2017automated,pitsilis2018detecting} have proposed methods for detecting offensive language. 
In this work, we refer to all of them as hate speech detection. 
We identified four categories of detection studies, based on the features and machine learning algorithms that they employ:
\textbf{Traditional Machine Learning:} 
Many works have proposed using traditional ML algorithms such as Support Vector Machines~\cite{davidson2017automated,salminen2018anatomy,frenda2019online}, Logistic Regression~\cite{saha2018hateminers}, etc. We characterized these studies based on the proposed features into \emph{content} and \emph{lexicon} based. 
\textbf{Content Based:} includes works that have obtained features solely from the text to detect hate speech, toxicity and offensive language, i.e., TF-IDF, n-grams, POS tags, BoWV etc.~\cite{saha2018hateminers,davidson2017automated,salminen2018anatomy,de2018hate}. 
\textbf{Lexicon Based:} 
Scholars have extensively used syntactic and semantic orientations of the existing lexicons for detecting hate speech because keyword-based approaches show high false positive rates, mostly due to ignoring the context~\cite{frenda2019online,vidgen2020detecting,capozzi2019computational,sanguinetti2018italian,basile2019semeval,tellez2018automated,vargas2021contextual,malmasi2017detecting,elsherief2018hate,wulczyn2017ex}. 
\textbf{Deep Neural Based:} 
With the advancements in the Deep Neural Networks, a majority of existing literature has used neural networks and deep neural networks to detect hate speech~\cite{yousaf2022deep,gamback2017using,alshamrani2021hate,chen2019use,zhang2018detecting,pitsilis2018detecting,agrawal2018deep,zhao2021comparative,doostmohammadi2020ghmerti,zimmerman2018improving,indurthi2019fermi} and toxic comments~\cite{georgakopoulos2018convolutional,elnaggar2018stop,srivastava2018identifying,vaidya2020empirical}.
Note that there are methods that use DNN based architectures together with the content, and lexical based features, which we list separately in the subsequent category of \emph{hybrid approaches}.
The methods categorized as Deep Neural based use CNN, LSTM, Transformers to detect hate speech.
CNNs have been used in a variety of works for hate speech detection tasks~\cite{ribeiro2019inf,winter2019know,kamble2018hate,rozental2019amobee,khan2020hateclassify,park2017one}. 
Sequential models such as LSTM have been effective in detecting hate speech on social media~\cite{badjatiya2017deep,raut2022hate,qian2018hierarchical,agarwal2021combating,li2022anti,vashistha2020online,modha2019ld,ousidhoum2019multilingual}. 
Following the success of transformer networks trained on large amount of data (or Large Language Models), pre-trained models such as BERT~\cite{devlin-etal-2019-bert}, GPT~\cite{gpt_paper} and RoBERTa~\cite{liu2019roberta} have been used successfully in 
hate speech detection~\cite{mozafari2020hate,alatawi2021detecting,saha2019hatemonitors,caselli2020hatebert,wullach-etal-2021-fight-fire,mozafari2019bert}. 

\textbf{Hybrid Approaches:} 
Previous works have effectively combined features derived from the content, lexicon and deep neural network, and several works have proposed hybrid approaches, using an ensemble of classifiers using different sets of features
~\cite{grondahl2018all,founta2019unified,malmasi2018challenges,del2017hate,gao2017detecting,salminen2020developing,mathew2019thou,chatzakou2017mean,paschalides2020mandola,almerekhi2020these,ziems2005racism,badjatiya2019stereotypical}. 
Grondahl et al.~\cite{grondahl2018all} evaluated four state-of-the-art hate speech detection models trained on different datasets and found that the models only work well within their respective trained datasets, failing to generalizing across datasets.

\textbf{Multi-modal Based: } Hate speech can span across multiple modalities such as images, videos, memes, etc. Hence it is imperative to detect this type of content~\cite{gomez2020exploring}. 
Works have used text and images together to detect hate speech~\cite{yang2019exploring,singh2017toward} text and socio-cultural information~\cite{vijayaraghavan2021interpretable} and memes~\cite{gomez2020exploring,kiela2020hateful,pramanick2021momenta,lippe2020multimodal,velioglu2020detecting}.

\textbf{Misinformation Detection} 
Scholarships have also extensively used ML techniques to detect misinformation/ fake news. 
We identified four categories, based on the features and the machine learning algorithms that they use: 
\textbf{Traditional Machine Learning:} Many works have proposed using traditional ML algorithms such as Random Forrest~\cite{singhal2021prevalence,zhou2020fake,frenda2019online}, Logistic Regression~\cite{della2018automatic}, etc. We characterized these studies based on the proposed features into \emph{content based}, \emph{propagation based}, \emph{hybrid} and \emph{crowd intelligence}. 
\textbf{Content Based:} features such as number of content words (nouns, verbs, adjective), and the presence and frequency of specific POS patterns, TD-IDF, have been employed, to detect misinformation, in~\cite{singhal2021prevalence,zhou2020fake,horne2017just,della2018automatic,hosseinimotlagh2018unsupervised,potthast2018stylometric,rashkin2017truth,volkova2018misleading,wang2017liar,felber2021constraint,hakak2021ensemble}.
\textbf{Propagation Based:} Different to methods extracting features from the textual content, there are multiple works using the propagation structure, or spreading networks of content on social media to detect misinformation~\cite{wu2018tracing,sampson2016leveraging}. These works leverage the peculiar traits in underlying network structure of the dissemination of misinformation~\cite{shu2020hierarchical}, classifying news propagation paths for early detection~\cite{liu2018early}, and identifying higher-order mutual attention paths in the propagation structures~\cite{mishra2020fake}.
Scholarships have also used user based features such as number of friends, followers etc, to identify misinformation/ fake news~\cite{singhal2021prevalence,shu2018understanding,shu2019role}.
\textbf{Hybrid Approaches:} Combining the features derived from both the content and propagation structure of content on social networks, several works have proposed hybrid approaches to detect misinformation~\cite{ruchansky2017csi,shu2019beyond,jin2017multimodal,karimi2018multi,helmstetter2018weakly}.
\textbf{Crowd Intelligence: } Wisdom of the crowd can be effectively used to detect misinformation. Multiple works have leveraged implicit crowd intelligence as the content spreads on a social media for early detection of misinformation~\cite{kim2018leveraging,ma2018rumor,tschiatschek2018fake,vo2018rise,jin2016news,qian2018neural,shu2019defend}.

\textbf{Deep Neural Networks:} Recent advent of Deep Learning algorithms have allowed researchers to build detection systems without feature engineering. Multiple works have used advanced deep learning techniques such as self-attention~\cite{aloshban2020act}, adversarial learning~\cite{ren2020adversarial}, graph neural networks~\cite{bian2020rumor,yuan2021improving}, recursive neural networks~\cite{ma2018rumor}, and other deep learning architectures~\cite{shu2019defend,liu2020fned,ruchansky2017csi,monti2019fake,wang2018eann,kaliyar2021fakebert,lu2020gcan,wu2018tracing} to detect misinformation. 
\textbf{Knowledge Based:} Another line of research uses Knowledge Graph 
built through relational knowledge extracted from the Open Web and evaluate the truthfulness of contents by verifying them against the ``gold'' Knowledge Base~\cite{hu2021compare,kou2022hc,cui2020deterrent,dun2021kan,pan2018content}. Scholarships have used a combination of Knowledge Base and deep neural networks to find misinformation~\cite{song2021temporally,mehta2022tackling,xu2022evidence,cui2020deterrent}. 
\textbf{Multi-modal Based:} The scope of detecting misinformation expands beyond just detecting text based misinformation but to identifying multi-modal misinformation. Researchers have proposed multiple methods to tackle this problem leveraging the relationship between different modalities of content on social media~\cite{singhal2022leveraging}, learning shared representations of text and images~\cite{khattar2019mvae}, learning the shared representations of audio and visual information~\cite{shang2021multimodal}, learning transferable multi-modal feature representations~\cite{wang2018eann}, and fusion based methods~\cite{wu2021multimodal,qi2019exploiting,song2021multimodal}.

\textbf{Research Gaps.} 
Bozarth and Budak~\cite{bozarth2020toward} evaluated five representative classifier architectures for misinformation detection, and found that the performance of detection systems vary across datasets, hence prompting a need for building comprehensive evaluation systems. 
Similarly, the evaluation of hate speech detection methods by~\cite{grondahl2018all} suggests that traditional machine learning based methods can be as good as Deep Neural Networks based methods, and the focus of researchers should be more on designing richer, robust datasets.
The sizes of existing datasets for hate speech detection range from 6K comments~\cite{del2017hate} to 100K tweets~\cite{founta2018large}.
However, Madukew et al.~\cite{madukwe2020data} identified multiple limitations with the existing datasets for hate speech detection: i) conflating class labels, ii) varying definitions of hate speech across manual annotations, and iii) class imbalance issues, thus highlighting the need for benchmark datasets.  
Fairness remains a key challenge when building these detection tools, as the underlying biases in the groundtruth dataset or lack of various dialects can reflect in the classification results~\cite{bolukbasi2016man,sap-etal-2019-risk}. 
Scholarships should focus on building standard benchmarks for evaluating hate speech detection systems across different domains, different languages, and across different nuances of hate speech such as implicit hate speech~\cite{yin2021towards,arango2019hate,monnar2022resources}. 
Misinformation detection methods are topic specific, e.g., they detect misinformation about presidential election or COVID-19, etc. However, misinformation about other topics are under-worked and under-studied. For example, Singhal et al.~\cite{singhal2021prevalence} found that misinformation regarding security and privacy threats are also prevalent on social media. 
Despite the promise of zero shot, cross-lingual language models, such as multilingual BERT~\cite{devlin-etal-2019-bert}, they are limited for cross-lingual hate speech detection~\cite{nozza2021exposing}.
Future works should focus on building efficient and scalable cross-lingual hate speech detection methods. Scholarships should focus on detecting domain independent multi-modal misinformation~\cite{abdali2022multi}, and generalizable multi-modal hate speech capturing the complexities of hate speech embedded in memes~\cite{jennifer2022feels}.

\subsection{Moderation Policies}
\label{mpp}
We identified six broad categories from our analysis of papers that are shown in Table~\ref{metrics-user-study} and Table~\ref{metrics-data-study} in the Section~\ref{app}: \textbf{Consumption of (fake/ misinformation) news: } 
Some research studies have tried to understand how the users consume the news online, and what methods are employed by social media to verify the information shared on these platforms. 
Geeng et al.~\cite{geeng2020social} focused on the effect of warning labels that were added on Twitter, Instagram, and Facebook on posts related to COVID-19 misinformation. 
They found that most of the survey takers had a positive attitude however, a majority of participants discovered or corrected misinformation by using other means, most commonly web searches. Zhang et al.~\cite{zhang2022shifting} found that most participants determined the credibility of news regarding COVID-19 using other heuristics such as web searchers. 
This research corroborates with other research, where authors found that people use multiple heuristics on and off social media to determine the credibility of information~\cite{geeng2020fake,jahanbakhsh2021exploring,bhuiyan2021nudgecred,sherman2021designing,urakami2022finding,haque2020combating,lu2020government,flintham2018falling,varanasi2022accost,lu2021positive}. 
\textbf{Research Gaps:} While these research studies give direction on how users investigate fake news and employ warnings, there are some research gaps that the research community should investigate. 
The study demographics concentrate in the US and had a large percentage of young, tech-savvy participants (18–25 years old) as shown in Table~\ref{metrics-user-study}. The study also uses textual data, such as articles and some images. 
Studies with more diverse participants, and also those which focus on specific demographics, e.g., certain age, gender, ethnicity can help our understanding of the acceptance of different content moderation policies in different communities.
Also, most of these works are simulation-based, where, for example, participants interacted with simulated web pages that show fake news. 
Users might show a different behavior, or use different approaches for investigating statements given in images, memes, and videos, therefore more observatory studies are needed to understand the problem. 

\textbf{Engagement of users: } Some studies investigated how users interacted with moderated postings and the consequences on engagement when certain communities or influencers are deplatformed and moved to stand-alone or fringe websites. 
Zannettou~\cite{zannettou2021won} performed a data driven study on the soft moderation interventions employed by Twitter. 
He found that tweets that have warning labels tend to have a higher user engagement. This was also corroborated by researchers in~\cite{papakyriakopoulos2022impact}. However,~\cite{ling2022learn,kim2020effects} found a contrasting results, where user engagement on content with warning labels on TikTok and YouTube was found to be less.
Mena~\cite{mena2020cleaning} conducted a user study to understand the effect of warning labels on the likelihood of sharing fake news on Facebook. 
He found that flagging fake news has a significant effect on users' sharing intentions; that is, users are less willing to share content with the labels. This was corroborated in~\cite{park2021experimental,yaqub2020effects,pennycook2020implied,jia2022understanding,seo2022if,epstein2021explanations}.
Chandrasekhran et al.~\cite{chandrasekharan2020quarantined} found that quarantine made it more difficult to recruit new members on r/TheRedPill and r/The\_Donald, however they find that the existing members hateful rhetoric remained the same. 
Similarly, Shen and Rose~\cite{shen2022tale} found that Reddit's quarantine was effective in decreasing the posting levels, however the toxicity of users remained the same.
Trujillo et al.~\cite{trujillo2022make} found that quarantine did in fact reduced the activity of problematic users, however it also caused an increase in toxicity and led users to share more polarized and less factual news. 
A similar result is seen in the data driven study done by works of~\cite{ribeiro2020does,chandrasekharan2017you,ali2021understanding,rauchfleisch2021deplatforming,thomas2021behavior} which studied the effectiveness of deplatforming. 
Works such as~\cite{jhaver2021evaluating,saleem2018aftermath} found that deplatforming significantly decreased posting level, user engagement and toxic rhetoric of the users. 
\textbf{Research Gaps:} 
Most user studies have a skewed demographic (shown in Table~\ref{metrics-user-study}), especially in terms of self-reported political views, as there were more liberals than conservatives, which could have affected their findings. 
More studies with diverse participants can fill the gap. Also, data-driven studies focusing on specific groups of users with different cultures and backgrounds can help understand the factors that affect user engagement in practice.
While works have studied the effect of deplatforming, they only study on a few platform, including Reddit, Gab, and Twitter which can be seen in Table~\ref{metrics-data-study}. 
Studies on other platforms, such as Facebook and Instagram, can provide more insights, as these platforms are more popular among different populations. 
Moreover, as content moderation policies on these platforms are constantly evolving, the impact of such changes on user engagement can be studied. Future scholarships could propose and study interventions that can be placed during sharing process, e.g., disabling sharing or displaying a splash warning screen on the shared content, similar to \emph{quarantine}. 

\textbf{Effectiveness: } Works have examined the effectiveness of both soft and hard moderation techniques on 
social media platforms. 
Some works investigated whether moderation can lead to users moving to less moderated platforms. 

\emph{Soft Moderation Interventions: } A 2018 Gallup survey found that more than 60\% of U.S. adults were less prone to sharing stories from sites that were clearly labeled as unreliable~\cite{lapowsky}. Saltz et al.~\cite{saltz2020encounters} found that participants had a different opinion regarding Facebook COVID-19 warning labels, some perceiving them necessary step to inform users whereas others saw them as politically biased and an act of censorship. 
Many studies~\cite{jia2022understanding,kirchner2020countering,mena2020cleaning,kaiser2021adapting} found that interstitial covers, labels and flagging decrease the perceived accuracy of COVID-19 misinformation and fake news on Twitter~\cite{sharevski2022misinformation} and Facebook~\cite{clayton2020real,mena2020cleaning}. 
Previous research has also found that correcting or debunking fake news can significantly decrease users' gullibility to the story~\cite{chan2017debunking,nyhan2020taking,park2021experimental,kim2020effects,seo2022if,yaqub2020effects,papakyriakopoulos2022impact,lu2020government}. 
Seo et al.~\cite{seo2019trust} investigated users perceptions when they were exposed to fact checking warning labels and machine learning generated warning labels. They found that users tend to trust fact checking warning labels more than machine learning generated warning labels. 
However, previous works also demonstrated some fortuitous consequences from the use of warning labels. Pennycook et al.~\cite{pennycook2020implied} found an \emph{implied truth} effect, where the posts that included misinformation but were not accompanied by a warning label were considered credible by the users. 
Studies found that there can be an unintended \emph{backfire effect}, where participants strengthen their support for the false political news that has a warning label or they distrust the source that fact checked it~\cite{gao2018label,grandhi2021crowd}. 
A few studies~\cite{shen2022tale,thomas2021behavior} found that Reddit's quarantine was ineffective and may also increase the polarization in political
spaces. 

\emph{Hard Moderation Interventions: } 
Chandrasekharan et al.~\cite{chandrasekharan2018internet} studied the Reddit comments that were removed by moderators to find macro, meso, and micro norms enforced to remove problematic content such as hateful content. 
Chandrasekhar et al.~\cite{chandrasekharan2017you} found that Reddit's ban on r/fatpeoplehate and r/CoonTown was effective, where users drastically decreased their hate speech usage. A similar result was seen in~\cite{saleem2018aftermath}.
Schoenebeck et al.~\cite{schoenebeck2021drawing} showed that users prefer that the platforms remove harassing content. Thomas et al.~\cite{thomas2022s} found that content creators felt platform policies and community guidelines were at least somewhat effective at keeping them safe from hate and harassment. In~\cite{munger2017tweetment}, the author found that when a user was blocked by a user who had a large number of followers, that user significantly reduced their use of a racist slur. However, Jhaver et al.~\cite{jhaver2018online} found that users who use blocklist on Twitter were not being adequately protected from harassment. Targets of online harassment expressed frustration with the lack of available support tools and the ineffectiveness of current hard moderation interventions of social media~\cite{blackwell2017classification,musgrave2022experiences,sambasivan2019they,doerfler2021m,jhaver2019did,xiao2022sensemaking}. 
Facebook, Twitter, Instagram, YouTube, and other platforms have all banned controversial influencers for spreading misinformation, conducting harassment, or violating other platform policies~\cite{jones-down,depl,depl-1,depl-2}. 
With these social media banning or fact checking posts, many right-wing individuals, citing censorship, are flocking to communities with fewer restrictions such as Parler, Gab, etc.~\cite{flock,flock1,parler-fact2,conservfav}. 
Scholars have extensively studied this migration and how it affects content moderation, and whether it increases or decreases hate speech, etc.
Jhaver et al.~\cite{jhaver2021evaluating} studied the effectiveness of permanent bans on Twitter of three influencers. They found that banning significantly reduced the number of conversations about all three individuals on Twitter and the toxicity levels of supporters declined. This finding has been corroborated by studies carried out in~\cite{ribeiro2020does,trujillo2022make}.
Some scholarships have examined the effects on users rhetoric and whether their followers discuss the influencers who were deplatformed by social media~\cite{ali2021understanding,rauchfleisch2021deplatforming,rogers2020deplatforming,chandrasekharan2020quarantined}.
They found a common result, that deplatforming significantly decreased the reach of the deplatformed users, however the hateful and toxic rhetoric increased. Kumarswamy~\cite{kumarswamy2022strict} studied the changes to Parler moderation strategies after it was taken offline by Apple, Google and Amazon Web Services. He found that the overall toxicity of the users decreased. 
\textbf{Research Gaps:} 
Scholarships should investigate if labeling every news article can decrease the \emph{implied truth} effect.
Further, the impact of changes in content moderation policies can be studied on different platforms. 
One possible avenue for future scholars, could be to understand if there is an \emph{echo chamber} effect on users, when they are given the option to toggle off content with soft moderated label. 

\textbf{Support: } 
Works have studied if users tend to support or disapprove the interventions (i.e., moderation) taken by social media. Through a user study, one research found that 45\% of Americans think that social media companies should have a role in moderating user-created content, however that same research also found that 41\% of Americans think that social media sites suppress free speech~\cite{free-speec}. 
Geeng et al.~\cite{geeng2020social} found that users had a positive outlook on moderation interventions. Gon{\c{c}}alves et al.~\cite{gonccalves2021common} found that algorithmic moderation is perceived more favorable then human moderation, while in contrast Lyons et al.~\cite{lyons2022s} found that human moderation was perceived more favorable than algorithmic moderation. 
Riedl et al.~\cite{riedl2021antecedents} found that users grouped by their age, their education level, and their opposition to censorship, supported social media content moderation intervention. Xiao et al.~\cite{xiao2022sensemaking} found that 
users want social media platforms to improve design and moderate content more proactively. 
Works such as~\cite{jhaver2019did,schoenebeck2021youth,saltz2020encounters,lu2020government,duffy2022platform,blunt2020posting} have found opposition to content moderation interventions, finding that users tend to echo that they are biased and are arbitrarily applied. 
Participants in~\cite{myers2018censored} echoed that there should be governmental regulation as they felt that it infringes their First Amendment right to free speech. 
\textbf{Research Gaps:} Most user studies focused on participants that were using mainstream social media websites. Future scholarships should have participants from fringe websites, to understand what kind of regulations would be most interesting to them.
Most of the user studies did not account for ethnicity as well as cultural differences. 

\textbf{Removal Comprehension: } A key aspect of understanding trust \& support in social media interventions is to understand if end-users can comprehend why their content was removed. 
Jhaver et al.~\cite{jhaver2019did} found that over a third of the participants did not understand why their content was removed and 29\% expressed frustration. Haimson et al.~\cite{haimson2021disproportionate} found that conservative users echoed the claim that their content was removed because they perceive social media platforms as heavily controlled by liberals, whereas black and transgender participants echoed that their content was removed because they were expressing their marginalized identities.
Schoenebeck et al.~\cite{schoenebeck2021youth} found that 41\% of the youth participants do not trust social media platforms, this can directly point to youths not comprehending why their content was removed.
Works such as~\cite{myers2018censored,vaccaro2020end,juneja2020through,gonccalves2021common,duffy2022platform} found that users complained about social media companies not disclosing the specifics as to why their content was removed. Scholars have also reflected upon the importance of transparency in content moderation systems
~\cite{jhaver2018online,seering2019moderator,suzor2019we,eslami2019user,juneja2020through}.
\textbf{Research Gaps:} Further research should investigate novel and effective designs for a redressal system, where users whose content is removed are given specific details as to why their content was taken down. 

\textbf{Fairness and Bias:} Social media platforms play a decisive role in promoting or constraining civil liberties~\cite{denardis2015internet}. 
How platforms make these decisions has important consequences for the communication rights of citizens and the shaping of our public discourse~\cite{costodian}. 
Shen and Rose~\cite{shen2019discourse} studied how the discourse on content moderation is polarized by users’ ideological viewpoints. They found that right-leaning users invoked censorship while left-leaning users highlighted inconsistencies in how content policies are applied. 
Works have studied users' opinions about bias and fairness of content moderation on various social media platforms. Lyons et al.~\cite{lyons2022s} found that users perceive human moderation as more fair and less biased than algorithmic moderation. However, Gon{\c{c}}alves et al.~\cite{gonccalves2021common} found that algorithmic moderation is perceived to be more transparent and less biased than human moderation.
Conservatives have often described the moderation decisions by Twitter, Facebook, etc., as biased and they claim that these companies censor their point of view~\cite{free-speec,myers2018censored,jhaver2019did,saltz2020encounters,schoenebeck2021drawing,suzor2019we,langvardt2017regulating,roberts2018digital,jahanbakhsh2021exploring}. 
Jhaver et al.~\cite{jhaver2018online} found that users who are on the Twitter blocklist feel they are blocked unnecessarily and unfairly. Roberts et al.~\cite{roberts2018digital} found that moderation is sometimes unfair for people in the marginalized communities.
Scholarships have also looked into the inconsistencies and unfairness of moderation decisions and the harm of moderation on marginalized communities~\cite{haimson2021disproportionate,seering2019moderator,vaccaro2020end,blunt2020posting,juneja2020through,gonccalves2021common,schoenebeck2021youth,costodian,musgrave2022experiences,duffy2022platform}.
\textbf{Research Gaps: } Scholarship should further investigate approaches for improving the precision and recall, as well as the algorithmic fairness of abuse detection systems. In addition, it is necessary to provide solutions for increasing the transparency of such algorithms, by for example, providing \emph{accuracy labels} on the warnings, or adding sources to factchecked claims. Moreover, since most findings on this topic are based on user studies, more data driven studies are required to investigate the validity of these findings. A more thorough discussion on increasing fairness is presented in Section~\ref{dis}. 

\section{Content Moderation in Practice} 
To answer our second research question, we studied how various social media platforms employ content moderation, what content do they moderate, and how the content moderation policies are defined, implemented, and enforced.

\textbf{Choice of Platforms:} 
We focused on the social media platforms that were investigated by prior research studies~\cite{geeng2020fake,sharevski2022misinformation,chandrasekharan2017bag} and from the recent events (i.e., the January 6 insurrection)~\cite{kumarswamy2022strict,mewe-fact,parler-fact2}. We chose fourteen diverse and popular platforms. These platforms include both mainstream (Facebook, YouTube, Instagram, TikTok, Snapchat, Twitter, Reddit, Twitch, and Tumblr) and fringe (4chan, Rumble, MeWe, Gab, and Parler) social media platforms, with different priorities, goals, and audiences. 

\subsection{Content Categories}
\label{cgcat}
\begin{figure}[t]
    \centering
    \includegraphics[width=0.85\columnwidth]{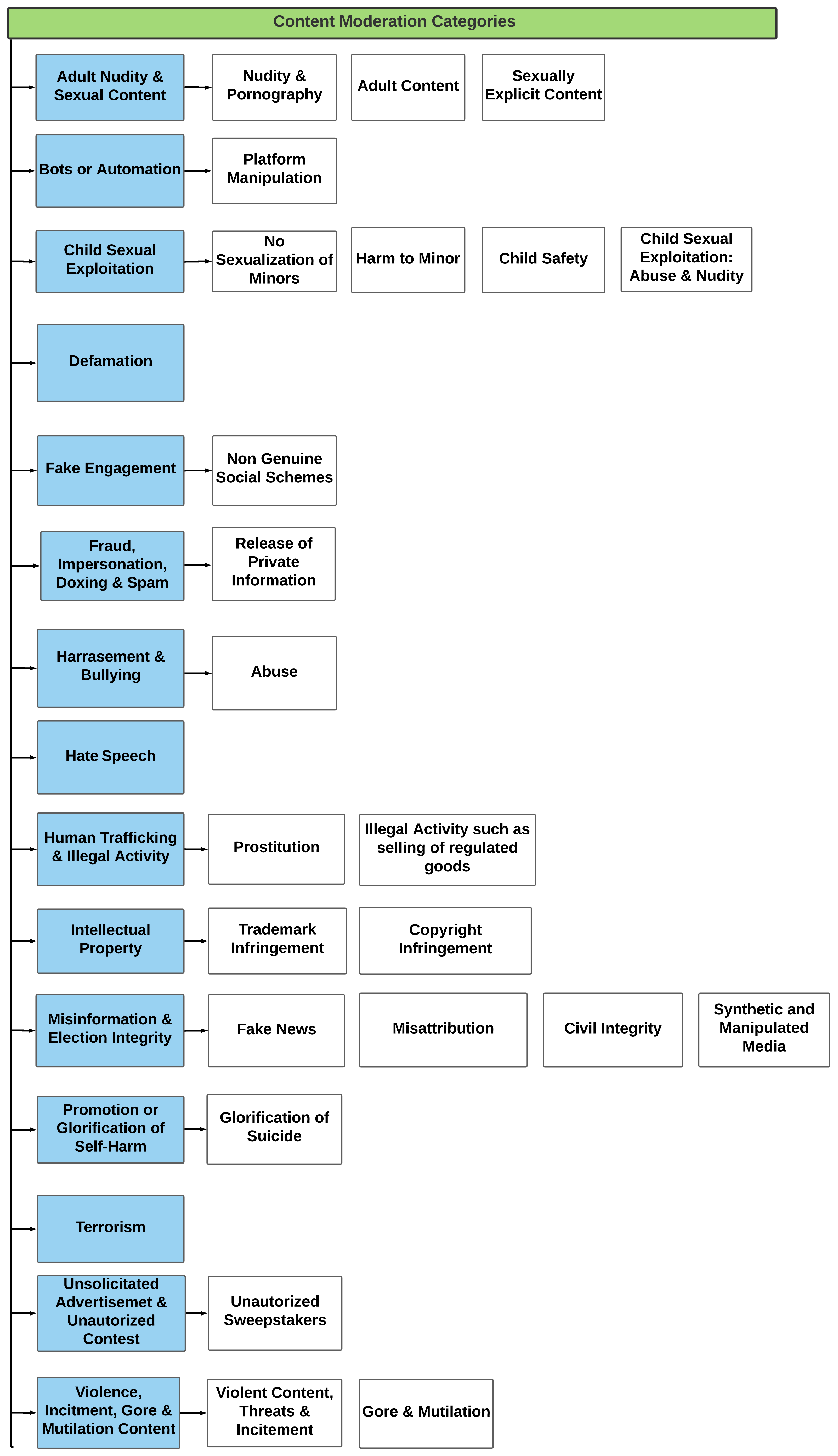}
    \caption{Content moderation categories based on analysis of guidelines}
    \label{fig:tmodcat}
\end{figure}
To better understand the prevalent content categories that are used by social media companies to regulate and moderate content, we studied each of the platforms'  \emph{Community Guidelines/Standards (CG/CS)} that were published on their \emph{US website} during June--August 2021. We investigated and compared the community guidelines of social media platforms for five countries, i.e., India, Germany, Australia, Brazil, and South Africa. Interestingly, we found no change in the guidelines, hence our analysis is not specific to the US. Table~\ref{guidell} in Section~\ref{anana} shows the results of this comparison. 
We also studied the CG of every platform in September 2022 to account for any substantial changes. We found a substantial change in the CG of Parler.\footnote{Example of changes in Parler CG: \url{https://tinyurl.com/yda6pfmj}}
Since each platform has a different nomenclature to how they describe some of the categories, we applied the open coding process~\cite{glaser2017discovery} to categorize them. 
One of the authors manually looked at all the social media platforms CG/CS to create categories until no new categories emerged. Then applying an iterative process new categories were added, or existing ones were reorganized. To create these categories, we followed certain guidelines: (1) Read through the CG/CS, and identify themes and sub-themes; and (2) While creating the categories, identify the meaning. Some of the categories were added based on majority score of where the category was placed by the social media platforms in their community guidelines. For example, \emph{Doxing} can also be part of \emph{Harrasement \& Bullying}, however, we found that a majority (12) platforms were placing it in \emph{Fraud}, hence we placed it in that category. 
Figure~\ref{fig:tmodcat} shows the content moderation categories. In total, we have fifteen main classes and twenty five subcategories in total. 
In the following, we define and describe each of the categories. 
The definitions of these categories were created based on differences in how various platforms were defining the categories, for example, Parler defines \emph{terrorism} as those groups officially recognized as such by the United States Government, whereas Twitter defines it's violent organization policy in a more detailed matter~\cite{parler-fight1,twitter-terror}.\footnote{Time-specific snapshot of platforms policies can be found at: \url{https://tinyurl.com/4wu9vzkc}}

\textbf{Adult Nudity \& Sexual Content}: 
Any consensually produced and distributed media that is pornographic or intended to cause sexual arousal, full or partial nudity, and simulated sexual acts. Exceptions include content related to artistic, medical, health, breastfeeding or education. 

\textbf{Bots or Automation}: 
Post content automatically, systematically, or programmatically, overutilize the service via automated tooling.

\textbf{Child Sexual Exploitation}:
Any type of abuse or sexual exploitation, i.e., nudity toward a child, any form of images, videos, text, or links that promote child sexual exploitation, sending sexually explicit media, trying to engage a child in a sexually explicit conversation.

\textbf{Defamation}:
Attacking in the form of oral or written communication of a false statement about another that unjustly harms their reputation.

\textbf{Fake Engagement}: 
Artificially increasing the number of views, likes, comments, or other metrics, selling or purchasing engagements, using or promoting third-party services or apps that claim to add engagement, trading, or coordinating to exchange engagement.

\textbf{Fraud, Impersonation, Doxing \& Spam}: 
Impersonation of individuals, groups, or organizations with intent or effect of misleading, confusing, or deceiving others, any fraudulent schemes, such as fake lotteries, phishing links, spamming users and comments, deceptive means to generate revenue or traffic, publishing of private information about an individual for malicious intent.

\textbf{Harassment \& Bullying}: 
Engage in the targeted harassment of someone, or incite other people to do so, sending threatening messages, and establishing malicious unsolicited contacts and threats.

\textbf{Hate Speech}: 
Direct attacks against people based on their protected characteristics, i.e., race, ethnicity, origin nationality, disability, religious belief, race, sexual orientation, gender, gender identity, and serious illness.

\textbf{Human Trafficking \& Illegal Activities}: 
Facilitate sex trafficking, other forms of human trafficking, or illegal prostitution, unlawful purpose or in furtherance of illegal activities including selling, buying, or facilitating transactions in illegal goods or services, as well as certain types of regulated goods or services.

\textbf{Intellectual Property}: 
Content that has the unauthorized use of a copyrighted image as a profile photo, allegations concerning the unauthorized use of a copyrighted video or image uploaded through our media hosting services.

\textbf{Misinformation \& Election Integrity}: 
Manipulating or interfering in elections or other civic processes. This includes posting or sharing content that may suppress participation or mislead people about when, where, or how to participate in a civic process, as well as sharing synthetic or manipulated media that is likely to cause harm.

\textbf{Promotion or Glorification of Self-Harm}:
Promotes suicide, self-harm, or is intended to shock or disgust users, content that urges or encourages others to: cut or injure themselves; embrace anorexia, bulimia, or other eating disorders; or commit suicide.

\textbf{Terrorism}: 
Promotes, encourages, or incites acts of terrorism, content that supports or celebrates terrorist organizations, their leaders, or associated violent activities.

\textbf{Unsolicited Advertisements \& Unauthorized Contests}: 
Unsolicited, unrelated advertisements, unauthorized contests, sweepstakes, and giveaways.

\textbf{Violence, Incitement, Gore \& Mutilation Content}:
Any threats of violence towards an individual or a group of people, content inciting people to commit violence, any media that depicts excessively graphic or gruesome content related to death, violence or severe physical harm, or violent content that is shared for sadistic purposes, severely injured or mutilated animals. There are however exceptions such as content may be made for documentary or educational content, religious sacrifice, food processing, and hunting.

\textbf{Discussion:} We found that interestingly some topics, such as \emph{misinformation}, have different definitions on each platform, and certain platforms give no definition of it. Our analysis also shows that in some categories, such as Adult Nudity \& Sexual Content \& Fake Engagement, there is no consensus among platforms on what content to moderate, which makes moderation arbitrary. Even platforms that are similar in nature (i.e., attract like-minded people), such as Parler and MeWe, do not prohibit hate speech content the same way. 
Therefore, there is a need for not only the computer scientists but also scholars from other disciplines such as psychology, social sciences, and law to come together and come to a consensus at least at the definition level. 
 
\subsection{Content Moderation Guidelines and Polices}
\label{cmp}
Table~\ref{moderation-1} shows the content moderation policies mentioned in platforms' CGs (soft or hard) as well as approaches used by them for enforcing these policies (human or ML). We also checked if any of these platforms claim to be doing factchecking. Table~\ref{moderation-1} shows \emph{No} if some policy is not defined, \emph{Yes} if it is defined, \emph{Partially} if it is only defined for some moderation categories mentioned in the previous section. Below we describe the findings based on Table~\ref{moderation-1}: 

\begin{table*}
\centering
\caption{Content moderation policies defined and enforced by different social media platforms }
\resizebox{\linewidth}{!}{%
\label{moderation-1}
\begin{tabular}{p{0.14\linewidth}|lllllllllllllll}
\hline
\multirow{2}{*}{\textbf{Moderation Strategy}}& & \multicolumn{11}{c}{\textbf{Platform}}\\ 
\cline{2-15}
& Parler & Gab & MeWe & 4chan & Rumble & Tumblr & Snapchat & TikTok & Reddit & YouTube & Facebook & Instagram & Twitter & Twitch \\ \hline
Factchecking & No & No &No&No&No&Yes&Partially&Yes&Yes&Yes&Yes&Yes&Yes&No \\
Soft Moderation & No & No &No&No&No&No&No&Yes&Yes&Yes&Yes&Yes&Yes& No\\
Hard Moderation & Partially&Partially&Partially&Yes&Yes&Yes&Yes&Yes&Yes&Yes&Yes&Yes&Yes&Yes\\
Human Moderators&Yes&X&Yes&Yes&Yes&Yes&Yes&Yes&Yes&Yes&Yes&Yes&Yes&Yes\\
ML Moderation & Yes&No&No&No&No&Yes&Yes&Yes&Yes&Yes&Yes&Yes&Yes&Yes\\
\hline 
\end{tabular}
}
\end{table*}
\begin{table*}
\centering
\caption{The content categories that are moderated on different social media platforms}
\resizebox{\linewidth}{!}{%
\label{moderation-2}
\begin{tabular}{p{0.35\linewidth}|lllllllllllllll}
\hline
\multirow{2}{*}{\textbf{Abusive Content Categories}} & \multicolumn{11}{c}{\textbf{Platform}}\\ 
\cline{2-15}
& Parler & Gab & MeWe & 4chan & Rumble & Tumblr & Snapchat & TikTok & Reddit & YouTube & Facebook & Instagram & Twitter& Twitch \\ \hline
Adult Nudity \& Sexual Content &Partially& Partially&X&Partially&X&Partially&Partially&X&Partially&Partially&Partially&Partially&Partially&Partially&\\
Bots or Automation& X & X &X&X&N/A&X&X&X&X&X&X&X&Allowed$^*$&X&\\
Child Sexual Exploitation & X&X&X&X&X&X&X&X&X&X&Partially&Partially&Partially&X&\\
Defamation &X&X &X&X&X&X&X&X&X&X&X&X&X&X& \\
Fake Engagement&N/A&X&N/A&N/A&N/A&X&N/A&X&X&X&X&X&X&X&\\
Fraud, Impersonation, Doxing \& Spam&X & X &X&X&X&X&X&X&X&X&X&X&X&X& \\
Harassment \& Bullying &N/A&N/A&X&N/A&N/A&X&X&X&X&X&X&X&X&X&\\
Hate Speech & Allowed$^{*}$&Allowed&X&Allowed&X&X&X&X&X&X&X&X&X&X&\\
Human Trafficking \& Illegal Activities&X & X &X&X&N/A&X&X&X&X&X&X&X&X&X& \\
Intellectual Property &X & X &X&X&X&X&X&X&X&X&X&X&X&X&\\
Misinformation \& Election Integrity& Allowed&Allowed&Allowed&Allowed&Allowed$^{*}$&X&X&X&X&X&X&X&X&X&\\
Promotion or Glorification of Self-Harm &N/A&N/A&N/A&N/A&X&X&X&X&X&X&X&X&X&X&\\
Terrorism & X & X &X&X&X&X&X&X&X&X&X&X&X&X&\\
Unsolicited Advertisements \& Unauthorized Contests &X & X &X&X&X&X&X&X&X&X&X&X&X&X&\\
Violence, Incitement, Gore \& Mutilation Content&Partially&X&X&X&X&X&X&X&X&Partially&Partially&Partially&Partially&X&\\

\hline 
\end{tabular}
}
\end{table*}

\textbf{Factchecking:} Parler, Gab, MeWe, Twitch, Rumble \& 4chan do not perform any factchecking on the content, also these social media platforms were created with a promise of less content moderation, as they call themselves \emph{champions of free speech}~\cite{parler-fact1,parler-fact2,gab-fact,mewe-fact}.
Snapchat performs factchecking only on advertisements, including political advertising~\cite{snap-fact}. Meanwhile, platforms like Tumblr, TikTok, Reddit, YouTube, Facebook, Instagram, \& Twitter factcheck claims that are posted on its platform. However, they do not specify if they fact check all the news. Based on the previous works though it seems these platforms only factcheck some topics, such as COVID-19 and elections.

\textbf{Hard Moderation:} Parler enforces some hard moderation but is kept to a minimum, as they mention ``We [Parler] prefer that removing users or user-provided content be kept to the absolute minimum''~\cite{parler-community}. 
After Parler was banned from Apple App Store \& Amazon following the January 6th Capitol roit~\cite{parler-ban}, Parler changed its community guidelines and is now not allowing \emph{hate speech} on iOS, but allowing on Android or the web~\cite{parler-come}. However, Parler does not remove so-called \emph{fighting words}, which are not protected as an exercise of the right to free speech~\cite{parler-fight1}. 
On Gab, users who are posting content from the ten categories described in Table~\ref{moderation-2}, would be subjected to a ban or their content will be taken down. MeWe also has a very similar approach to that of Gab, however, MeWe does not allow hate speech on its platform. 4chan, Tumblr, Snapchat, TikTok, Reddit, YouTube, Facebook, Instagram, Twitter, Rumble \& Twitch have more stringent hard moderation rules. Users can expect to have their content taken down, shadow banned, temporarily banned, or permanently banned from the platform. 

\textbf{Discussion:} Our analysis shows that there is no universal method for hard moderation across all social media platforms, and each platform follows a different approach, which can be due to the limitations that each of these methods has. Therefore, even more, rigid analysis is needed to measure and compare the performance of these methods and also propose new models that can improve their performance. 

\textbf{ML or Human-based moderation:} Parler, Tumblr, Snapchat, TikTok, Reddit, YouTube, Facebook, Instagram, Twitter, \& Twitch have both humans and AI as moderators. 
We were not able to find information about what kind of moderators are employed by Gab. 4chan has human moderators and \emph{janitors}, who are volunteers that can remove threads or replies on the imageboard they are assigned to, they can submit a request for a ban or warning of a user to the moderators (they themselves cannot ban the users). MeWe and Rumble also have human moderators~\cite{mewe-fact,rumble-mod}. 

\textbf{Soft Moderation:} We found that Parler, Gab, MeWe, 4chan, Tumblr, Rumble, Snapchat \& Twitch do not perform soft moderation on the content that is posted on their respective platforms. TikTok has started putting warning labels on the content where the facts are inconclusive or content is not able to be confirmed, especially during unfolding events. However, they have not specified what is considered as ``unfolding'' events. 
When a viewer tries to share a video that is flagged, they would be shown that the content is flagged for unverified content, and they can either cancel or share the post. 
Reddit's soft moderation technique is \emph{quarantine}. The main purpose of quarantine is to prevent users from accidentally viewing the content. Reddit also gives an explanation of why the community is in quarantine. However, a user can still view the content by clicking on the continue button. Twitter also has the same mechanism both for posts and also for accounts, a user can however still view the content~\cite{twitter-label}.
YouTube, Facebook, Instagram \& Twitter also have a very similar soft moderation policy. 

\textbf{Gaps in Practice:} 
Future scholarships should investigate ways where mainstream social media can also safeguard the First Amendment rights. Also, in practice, soft moderation has only been used for labeling misinformation. However, soft moderation might also be useful in labeling other types of abuse, e.g., malicious content, phishing urls, bot generated content, hate speech, etc. 

\subsection{What Content Category Gets Moderation?} 
We extracted the topics that social media platforms claim to be moderating in their community guidelines. 
Table~\ref{moderation-2} shows all different types of abusive content that each of the platforms have mentioned in their CGs. Two coders then labelled the categories for each platform. Coders followed the given nomenclature to label the categories: 
\emph{X} if the content is not allowed, \emph{Partially} if it is only allowed for some sub-topics, e.g., only for specific types of nudity and not for porn, \emph{N/A} if the guidelines do not mention that topic, \emph{Allowed} if the topic is allowed in the platform, and \emph{Allowed$^*$} if the topic is allowed, however, it has restrictions. Using this methodology, we calculated the inter-coder reliability score. We found a substantial agreement of 0.76. 

\textbf{Misinformation \& Election Integrity:} We found that any content that is posting misinformation about COVID-19 or elections are not allowed on Tumblr, TikTok, Reddit, YouTube, Facebook, Instagram, Twitter, \& Twitch. However, Parler, Gab, MeWe, Rumble, \& 4chan allows misinformation to be present on its platform and they do not moderate them.

\textbf{Hate Speech:} Hate speech is not allowed on MeWe, Tumblr, Snapchat, TikTok, Reddit, YouTube, Facebook, Instagram, Rumble Twitter, \& Twitch. Parler does not allow any content that is hateful on iOS devices, but it still allows it on other devices~\cite{parler-come}. Gab and 4chan do not remove hate speech on their  platform~\cite{gab-misi,zelenkauskaite2021shades}. 

\textbf{Adult Nudity \& Sexual Content:}We found that adult nudity \& sexual content of any kind are not allowed on MeWe, and TikTok. In Parler, users are allowed to post images, videos, depictions, or descriptions of adult nudity or sex as long as they are designated as \emph{sensitive} (NSFW)~\cite{parler-fight1}. However, if a user posts any content that is containing nudity or sexual content, and the user has not designated it as NSFW, then that content is removed. Exceptions are made for spiritual artwork or posts by a verified art gallery. 
In 4chan, users are allowed to post Anthropomorphic pornography, Grotesque images, and Loli/shota pornography only in /b/ board~\cite{4chan-rules}. Platforms such as Gab, Tumblr, Snapchat, Reddit, YouTube, Facebook, Instagram, Twitter, \& Twitch allow nudity content e.g. as a form of protest or for educational/medical reasons, with Twitch allowing individuals actively breastfeeding a child on stream. Twitter, Instagram, \& Facebook would apply a label to content involving breastfeeding, and images/videos shared in medical or health context.

\textbf{Bullying \& Harassment:} We found that any content that is bullying \& harassing a user, or a group is not allowed on MeWe, Tumblr, Snapchat, TikTok, Reddit, YouTube, Facebook, Instagram, Twitter, \& Twitch. Parler, Rumble, Gab, \& 4chan do not have a policy outlining this category.

\textbf{Promotion or Glorification of Self-Harm:} We found that promotion or glorification of self harm is not allowed on Tumblr, Snapchat, TikTok, Reddit, YouTube, Facebook, Instagram, Rumble, Twitter, \& Twitch. Parler, Gab, MeWe, \& 4chan do not have a policy outlining this category.

\textbf{Fake Engagement:} We found that fake engagement is not allowed on Gab, Tumblr, TikTok, Reddit, YouTube, Facebook, Instagram, Twitter, \& Twitch. Parler, MeWe, Rumble, Snapchat, \& 4chan do not have a policy on this category.

\textbf{Defamation:} Defamation is not allowed across all the platforms. 

\textbf{Violence, Incitement, Gore \& Mutilation Content:} We found that some forms of Gore \& Mutilation content is allowed on Parler, YouTube, Facebook, Instagram, \& Twitter. On Parler, users should designate content as NSFW, if they fail, then the content will be removed. YouTube, Facebook, Instagram, \& Twitter have exceptions for religious sacrifice, food preparation or processing, and hunting. 

\textbf{Child Sexual Exploitation:} 
We found that Facebook, Instagram, \& Twitter, allow only educational content for child sexual exploitation i.e., documentaries, news media reportage. These types of content are normally accompanied by a label in Facebook and Instagram.
While platforms explicitly do not mention how they identify images or videos depicting child sexual exploitation, previous works have reported that major social media companies use PhotoDNA~\cite{gorwa2020algorithmic}. 

\textbf{Fraud, Impersonation, Doxing \& Spam:} We found that across all the platforms, any type of fraud, impersonation, doxing \& spam are not allowed. 

\textbf{Terrorism:} We found that across all the platforms, the promotion of terrorist propaganda, or violent extremism are not allowed. This includes recruiting for a violent organization, or using the insignia or symbol of violent organizations to promote them. Previous work has reported that various social media companies use a \emph{Shared Industry Hashed Database}. 

\textbf{Bots or Automation:} All the platforms except Twitter \& Rumble prohibit the use of bots or automation to post content or over-utilize the service by sending excessive queries. Twitter allows users to send automated tweets, sending replies and mentions etc., however, they have to authorize an app or service via OAuth~\cite{twitter-bots}. We were not able to find any such policy outlined in Rumble's CGs.

\textbf{Unsolicited Advertisements \& Unauthorized Contests:} None of the platforms allow users to post unsolicited advertisements or hold any unauthorized contests. Tumblr has a separate policy outlying the rules to hold contest, sweepstakes, and giveaways~\cite{tumblr-contest}.
While platforms have regulations for product placements and influencers adverts, it is currently out of the scope of this paper.

\textbf{Intellectual Property:} All the platforms ban any content that infringes the copyright. Any content that infringes the copyright will be removed. Content ID is a state of the art system that is used to detect copyright infringement content~\cite{soha2016monetizing}. However, content that is satire, parody, and news reporting among others are not in violation of intellectual property, and countries such as the US and some other countries, follow the \emph{fair use} doctrine, whereas the EU has some exceptions~\cite{guibault2007study,Copyright1}.

\textbf{Human Trafficking \& Illegal Activities:}
All the platforms except Rumble do not allow any content about Human Trafficking \& Illegal Activities. We were not able to find any such policy outlined in Rumble's community guidelines. 

\textbf{Discussion: } We found that content that is legally required to be removed from the social media platforms, such as \emph{Child Sexual Exploitation}, \emph{Violence}, \emph{Intellectual Property}, and  \emph{Terrorism}, has a more uniform moderation across all the fourteen platforms. We also found a very consistent pattern in fringe social media platforms with respect to \emph{Misinformation \& Election Integrity}, where all these platforms allowed them.
We also see that there are spectrum of differences in the definitions of \emph{Adult Nudity \& Sexual Content} on Parler and 4chan, both allowing them, but they have various differences to what is allowed, however, mainstream social medias such as Twitter, Facebook have uniformity in the content that may be allowed.

\subsection{Community Guidelines Comprehensibility} 
\begin{figure}[t]
    \centering
    \includegraphics[width=0.8\columnwidth]{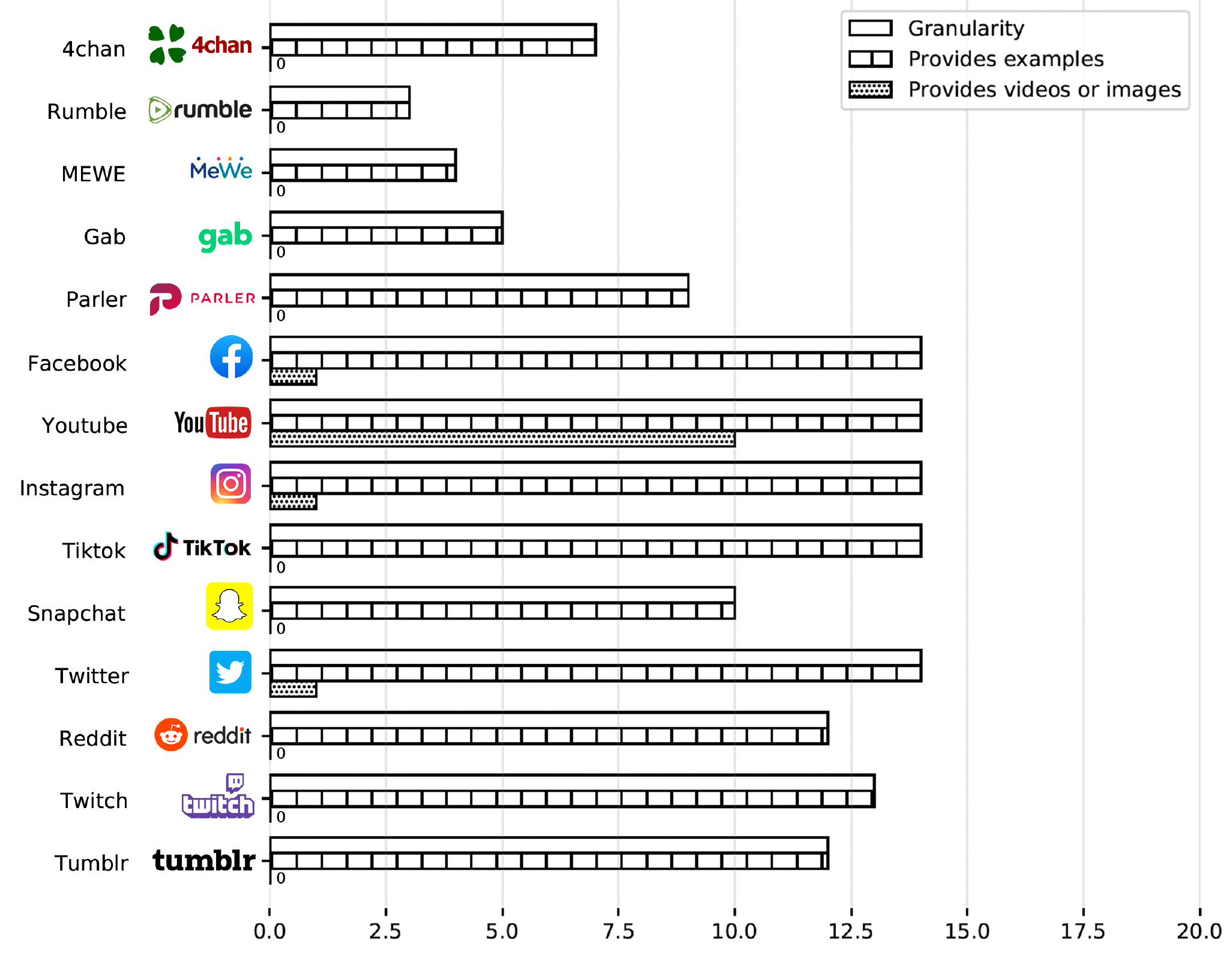}
    \caption{Community Guidelines Comprehensibility }
    \label{fig:moderation}
\end{figure}

Comprehensibility of community guidelines means whether the end-user can understand completely what type of content is allowed on the platforms and what would be the repercussions if there is a violation of these terms. The goal of this analysis is to identify places where guidelines are not comprehensible enough. For that, we first check the granularity of community guidelines in terms of covering various content categories, and then we check if they have provided examples, images, and videos for each of the categories.  
We used the same mechanism to label the categories as described in Section~\ref{cmp}. Figure~\ref{fig:moderation} shows the results. It shows three broad categories:  (1) granularity, (2) provides an example, \& (3) provides videos or images. Two authors independently coded all the categories. For inconsistent results, coders discussed how to resolve disagreements. To assess the inter-coder reliability, we performed a Cohen-Kappa test~\cite{schuster2004note}. The Kappa score was 0.75, which shows substantial agreement.   

\textbf{Granularity:} Granularity can help users to understand in more detail the content that is allowed and what content is not allowed. To measure granularity, we used the data provided in Table~\ref{moderation-2} and coded \emph{Yes} and \emph{Partially} as 1 and \emph{No} as 0, and then computed the sum. 
In our analysis, we found that YouTube, Facebook, TikTok, Instagram, \& Twitter were the top five platforms to have very granular CGs. However, we found that MeWe and Rumble were the only two platforms, whose community guidelines were not granular, except for four and three sub-categories, respectively. The average number of categories for which the platforms' community guidelines were granular was ten out of fifteen categories, the minimum was four, and the maximum was fourteen. 

\textbf{Provides Examples:} In order to facilitate the rules, social media companies also provide examples to help users understand what type of content will not be permissible and what will be. In our analysis, we found that YouTube, Facebook, TikTok, Instagram, \& Twitter were the top five platforms that were providing examples in their community guidelines. Rumble and MeWe were the two platforms with the least number of examples in the subcategories studied (i.e., three \& four respectively). 
The average number of categories for which the platforms were providing examples was ten out of fifteen, with the minimum as four, and the maximum as fourteen. 
Interestingly, among the fringe social media that we studied, Parler provided the most number of examples, i.e., nine.
For the subcategory, \emph{Violence, Incitement, Gore \& Mutilation Content} we found that 4chan, \& Snapchat were only providing examples for violence and incitement, however, these platforms failed to provide examples for gore and mutilation content. TikTok provides examples for automation, however, the platform does not provide examples about the use of bots. Twitch provides an ample number of examples for misinformation, however, the community guidelines do not mention posts/videos challenging the integrity of an election.

\textbf{Provides Videos or Images:} Audio and visual context aid users to grasp the content that will not be moderated. We found that only YouTube provides videos for the ten sub-categories. However, we found that for misinformation and election integrity, Facebook, Instagram, \& Twitter with YouTube provided visual material for the user. The average number of categories for which the platforms' community guidelines were providing videos or images was one out of fifteen, with the minimum as zero, and the maximum as ten. 

\textbf{Summary:} Figure~\ref{fig:moderation} shows that only Youtube community guidelines provide videos or images. We also found that a lot of mainstream social media, such as Tumblr, Twitch, Reddit, etc., do not provide such. Future scholarships can investigate if by providing such examples, users would pay more attention to the guidelines and whether it helps users to understand them. This can have implications regarding users' complaints that they do not understand the reason behind the removal of their content. 
\section{Discussion}
\label{dis}

\textbf{Content Moderation from the Legal Perspective.} 
Freedom of speech is equivocally a basic human right, and prior works have found that users complain that their voices are being suppressed by social media companies~\cite{myers2018censored,jhaver2019did,saltz2020encounters,schoenebeck2021drawing,suzor2019we}. 
Gillespie~\cite{gillespie2010politics} argued that platforms are innately political entities that have attempted to maintain an image of neutrality. It is this image of neutrality in the domain of speech that has come under question most in recent years, and the inability of platforms to be truly neutral is core~\cite{seering2020reconsidering}. 
People expect to have the same protections on social media platforms as they would have in the real world, while prior works have found that users want government oversight~\cite{myers2018censored,chung2022social} and a possible solution to this is to have government actors determine which platforms' content moderation practices could be subject to government oversight, with the platforms following legally defined set of rules. 
Platforms have to follow laws in the jurisdictions in which they are operating. For example, in the US, social media companies are not considered as a \emph{state} entity under \emph{Marsh v. Alabama}, and hence do not have to guarantee First Amendment protection to the user for protected speech~\cite{1946marsh}. Social media platforms are also shielded by Section 230 of the Communication Decency Act (CDA) which gives online intermediaries broad immunity from liability for user-generated content posted on their sites~\cite{section230def}. In the landmark case of \emph{Zeran v. America Online, Inc.}, the purpose of \S 230 is to encourage platforms to act as a \emph{Good Samaritans} and to take an active role in removing offensive content~\cite{1997zeran}. However, two currently pending cases in the Supreme Court, i.e., Gonzalez v. Google LLC~\cite{2021gonzalez} \& Twitter, Inc. v. Taamneh~\cite{2022twitter} could upend \S 230 and consequentially pose serious risks to Internet speech~\cite{goldman2019section,klonick2017new}.
In India, pursuant to Article 4(d) of India's Information Technology (Intermediary Guidelines and Digital Media Ethics Code) Rules, 2021, social media platform companies must publish a monthly report regarding their handling of complaints from users in India, including actions taken on them~\cite{etchiccs-code,etchiccs-code1}. 
In Germany, The Network Enforcement Act (NetzDG) obliges social media platforms with over 2 million users to remove \emph{clearly illegal} content within 24 hours and all illegal content within 7 days of it being posted or face a maximum fine of 50 million Euros~\cite{netdgz}. While the law is one of the most stringent in the world, scholarships, and human rights organizations have criticized it for incentivizing social media platforms to preemptively censor valid and lawful expression~\cite{gerhum}. 
In Australia, users can submit a complaint to the Australian eSafety Commissioner and get a remedy from the commissioner by using the available safety tools and resources~\cite{esafe}.

One way to moderate content in the grey areas and still uphold the statutory and legal doctrines is to apply soft moderation tags to posts that are inaccurate or explicit. This would be a welcoming step for users who echo that platforms are deliberately censoring their point of view~\cite{haimson2021disproportionate,myers2018censored}.

\textbf{Transparency in Moderation.}  
 Over the past few years, researchers have reflected on the importance of transparency in content moderation~\cite{seering2019moderator,suzor2018evaluating,suzor2019we,myers2018censored,schoenebeck2021youth,jhaver2019did}. Legal scholars have argued that platforms should disclose the content moderation policies and procedures and in order to increase users' trust~\cite{langvardt2017regulating,wilson2020hate}. Prior works have also found that users often do not understand why their content was removed, as they get a very generic response from the social media companies~\cite{myers2018censored,schoenebeck2021youth,jhaver2019did}. 
We concur with the suggestion from D{\'\i}az and Hecht~\cite{diaz2021double}, that social media platforms should provide transparency reports which will be an essential step in allowing users and watchdog organizations alike to identify issues ranging from lack of language expertise to biased algorithms. 
The transparency report should report the identities of organizational trusted ﬂaggers and the distinct content policies for which they have special ﬂagging privileges, the types of automated tools deployed to identify and remove the offending content, such as the use of hashing systems and natural language processing systems and whether the moderators are employed by the company itself or are they outsourced to a third party. We also echo that platforms should make public log reports for the moderated content and more social media companies should provide APIs for independent researchers to review the mechanisms and also conduct a fairness analysis on the system.
Platforms should also publish the number of orders received from government agencies to remove content or suspend accounts, and whether the platform took action, and if so was it based on actual infringements of the law or was it based on the violations of community guidelines? 

\textbf{One Size does not Fit All.}  
The amorphous nature of social media platforms’ content moderation policies gives companies enormous discretion in their enforcement, however, they should exercise this discretion in a way that truly balances free expression with equity and prioritizes user safety and due process~\cite{diaz2021double}. 
Following the death of George Floyd, users have tried to make their voices heard both in offline and online spaces. However, due to companies enforcement for speech talking about terrorism or racism, users especially those of color have found that their voices are getting suppressed~\cite{race,race1}. 
Prior researches have also corroborated that, voices of marginalized communities are often suppressed~\cite{haimson2021disproportionate,roberts2018digital,seering2019moderator}.
On the other hand, hate speech policies attempt to take a more measured approach with narrower restrictions, which can create a convoluted order of operations that ends up protecting powerful groups or allowing all but the most explicit attacks based on protected characteristics~\cite{diaz2021double}. Hence, platforms should take steps to redress their moderation systems, so that all users are treated equally. 
We recommend that social media platforms should take into account the current political and cultural context of the country when they are moderating content. This will help in users from marginalized communities the freedom of expression and at the same time punish the powerful groups that are often involved in incitement of violence or predominately engaging in hateful rhetoric. 

Demonetization of videos is one of the broader suites of content moderation governance mechanisms that is available to YouTube~\cite{gorwa2019platform}. However, creators have echoed that inconsistencies present in the demonetization process lead to beliefs of being algorithmically controlled or censored~\cite{caplan2020tiered}. Interestingly, it has been found that YouTube was treating established media personalities differently than they were treating users with a few thousand subscribers~\cite{youtube-modersep,kumar2019algorithmic,dunna2022paying}. 
Hence, we echo that content moderation policy must be overhauled to account for power flux among different groups such as users with large follower bases, i.e., influencers, and structural inequalities that may curtain opportunity for equal access to platforms. 
Platforms should acknowledge the fact that the ability to protect the speech of ordinary people when challenging the elite or influential people of the society requires a different matrix than the protections necessary for inﬂuential ﬁgures who often have multiple avenues for disseminating their message. Platforms should publish policies that govern public figures, heads of state, and other influencers in community guidelines.
Social media companies should make sure that both the automated moderation tool and human moderators are able to accurately assess different languages, dialects, slang, and related variations of context. 
Social media companies should make sure that their AI systems are not potentially discriminating because of the train cases given to them. 

\textbf{Collaborative Human-AI Decision Making.} 
With the growing scale of content, platforms employ human moderators, ML-based moderators, or both to regulate the content. While human moderation comes at a cost of time and scalability, ML-based moderation tools also suffer from the lack of training examples specific to regions or local dialects. It is arguably impossible to make perfect automated moderation systems because their judgments need to account for the context, the complexity of language, and emerging forms of obscenity and harassment, and they exist in adversarial settings where they are vulnerable to exploitation by bad actors~\cite{jhaver2019human,mcdaniel2016machine}. Even though hard coding the criteria helps in scalability and consistency, they still suffer from being insensitive to the individual differences of content, for example, when distinguishing hate speech from newsworthiness~\cite{caplan2018content}. Hence, there is a need for more proactive human interventions to offset the errors done by ML based moderation. However, more studies should be conducted to examine the effectiveness of having interconnected moderation systems, where human moderators proactively provide feedback on ML-based moderation. 

Fairness is another challenge of such systems, especially when they mainly rely on user-labeled data. Therefore, it is also crucial to develop fully or partially automated methods and algorithms that minimize the impact of bias on moderation decisions. 
One more human-based method for increasing fairness of content moderation is allowing users to appeal decisions, where an impartial, independent authority can verify the decisions. An example of this type of impartial jury is \emph{Oversight Board}~\cite{oversi}. 
The Oversight Board appeals process gives people a way to challenge content decisions on Facebook or Instagram. While this is a good step forward, the oversight board does not evaluate all the submitted cases but selects eligible cases that are difficult, significant, and globally relevant that can inform future policy~\cite{oversi1}. Pan et al.~\cite{pan2022comparing} work also found strong evidence of support for independent expert juries, similar to The Oversight Board. 

\textbf{Open sourcing vs. restricting access to content.} To provide a sense of fairness and accountability, open sourcing public data and public moderation logs are necessary. As we found in Section~\ref{lit}, previous works have largely studied moderation on platforms such as Twitter and Reddit, as they provide mechanisms for researchers to archive data. However, there are platforms from mainstream to fringe that are still restricting researchers from obtaining the vast amount of public data. 

\textbf{Uniformity vs. favorability. } With the Russian invasion of Ukraine, big tech companies such as Twitter, Facebook and Instagram, loosened their hate speech, for Ukranians to post about calling for general violence against Russians~\cite{ukrainreut}. While, this was ultimately changed three days later, such was not the case for protesters in Iran, where posts that include \emph{death to the dictator} - a key protest slogan, are being proactively flagged and taken down~\cite{wpiran,slate-irn,inter-irn,georgetown}. One has to ask, if platforms should remain neutral and uniform in implementing content moderation policies? 
In addition, more studies should be conducted to research the implications of using a threshold to detect hate speech more or less rigorously depending on the offline events. 
\section{Limitations \& Future Work}
In this work, we focus on the platforms from the consumer point of view and not from business point of view. We plan to study that in the future to look at the platforms from a business point of view.

\section{Conclusion}
In this work, we have presented three taxonomies based on an extensive review of the social media community guidelines and previous works. Using the taxonomies we answer the two research questions. We concluded that the most popular and mainstream social media platforms moderate for all categories studied as well as using both hard moderation and soft moderation categories. 
Out of these six platforms, only YouTube provides image or video examples. On the other hand, fringe platforms do not moderate for all the studied categories and prefer minimal interventions. 

\section{Acknowledgment}
This material is based upon work supported by the National Science Foundation under Award NSF III Medium 2107296 at the University of Texas at Arlington, and Awards CNS-1942610 and CNS-2114407 at Boston University. 
We also would like to thank professor Andrew Sellars for his helpful feedback and fruitful discussion on the legal aspect of content moderation. We would also like to thank our shepherd Dr. Jason (Minhui) Xui and anonymous reviewers their for helpful comments.

\bibliographystyle{plain}

\appendices
\begin{table*}
\centering
\caption{Community guidelines analysis of different countries. Note that \emph{\checkmark} denotes that the guidelines were similar to USA and \emph{\texttimes} that the platform is banned in the country studied.}
\begin{tabular}{c|c|c|c|c|c}
\hline
\textbf{Platforms}&\textbf{Germany}&\textbf{India}&\textbf{Australia}&\textbf{Brazil}&\textbf{South Africa}  \\
     \hline
    Parler & \checkmark& \checkmark&\checkmark &\checkmark &\checkmark\\ 
    4chan&\checkmark& \checkmark&\checkmark &\checkmark &\checkmark\\
    Gab&\checkmark& \checkmark&\checkmark &\checkmark &\checkmark\\
    MeWe&\checkmark& \checkmark&\checkmark &\checkmark &\checkmark\\ 
    Rumble&\checkmark& \checkmark&\checkmark &\checkmark &\checkmark\\ 
    TikTok&\checkmark& \texttimes$^{*}$&\checkmark &\checkmark &\checkmark\\ 
    Twitch&\checkmark& \checkmark&\checkmark &\checkmark &\checkmark\\
    Tumblr&\checkmark& \checkmark&\checkmark &\checkmark &\checkmark\\
    Reddit&\checkmark& \checkmark&\checkmark &\checkmark &\checkmark\\
    Snapchat&\checkmark& \checkmark&\checkmark &\checkmark &\checkmark\\
    YouTube&\checkmark& \checkmark&\checkmark &\checkmark &\checkmark\\
    Twitter&\checkmark& \checkmark&\checkmark &\checkmark &\checkmark\\
    Facebook&\checkmark& \checkmark&\checkmark &\checkmark &\checkmark\\
    Instagram&\checkmark& \checkmark&\checkmark &\checkmark &\checkmark\\
    \hline
    \end{tabular}
    \label{guidell}
\end{table*}
\section{Analysis of Community Guidelines of other countries:}
\label{anana}

We investigated and analyzed community guidelines of five countries, with the highest populations in each continent i.e., India, Germany, Australia, Brazil, and South Africa and used a VPN location, to access the community guidelines for the platforms. We excluded China because most of the platforms investigated in our paper are prohibited in China, and Chinese users rely on VPNs to access these websites, causing IP addresses to be inaccurate. Interestingly, we found no change in the guidelines. Table~\ref{guidell} shows them in detail. TikTok was banned in India in 2020, due to a geopolitical dispute with China~\cite{tikban-1}.

\section{Related work conference list}
\label{ekk}
\begin{table}[h]
    \centering
    \caption{Distribution of paper analyzed by the authors}
    \begin{tabular}{l|c}
    \hline
     \textbf{Conference}&\textbf{Number of papers}  \\
     \hline
    IEEE S\&P & 2\\ 
    ACM CCS & 1\\ 
    NDSS & 0 \\ 
    USENIX Security & 1 \\ 
    IEEE Euro S\&P & 0 \\ 
    Proc. of ACM CSCW & 26  \\ 
    ACM CHI & 25\\ 
    WWW & 14\\ 
    ICWSM & 16 \\ 
    AAAI AI & 6 \\ 
    ACL Anthology & 37 \\ 
    ACM WebSci & 6\\ 
    NeurIPS&3 \\ 
    IEEE/CVF& 2\\ 
    ACM KDD&5\\ 
    ACM WSDM & 4\\ 
    ACM CIKM & 2\\ 
    Other Conf. \& Workshops & 35 \\ 
    New Media \& Society & 6\\ 
    Political Behaviour& 5\\ 
    Other Journals& 52\\ 
    Law Reviews & 7 \\ 
    Thesis & 1 \\ 
    Books & 4 \\ 
    Transactions & 4 \\ 
    Others & 9 \\ \hline
    \end{tabular}

    \label{conf-list}
\end{table}
Table~\ref{conf-list} shows the distribution of papers from various conferences and journals that were analyzed by the authors. 

\section{Related work paper list}
\label{app}

Table~\ref{metrics-user-study} shows the paper that conducted user studies to understand moderation. In the table, we show the category of paper that was created in Section~\ref{lit}, we also show which platform the authors conducted their study on, the sample size, the demographics of the study and they type of the study conducted. Note that in sample size, if there is a number in red color brackets, that means that the authors of the study conducted multiple studies. We however, present the demographics of the study with highest number of participants. 

Table~\ref{metrics-detection} shows the papers that are about hate speech and misinformation detection. Table~\ref{metrics-data-study} shows the papers that are data driven study to unpack the intricate details about content moderation. Similar to Table~\ref{metrics-user-study}, we characterize the papers into the category, the platforms they studied. In this table, we also give the size of the dataset, if the dataset is public, whether the authors created the dataset or they used some other papers dataset, the type of data used and the intervention studied. The paper that are surrounded by red brackets, signifies that the authors studied both hate speech and misinformation moderation, whereas, the paper that is given in blue color signifies that that paper also conducted a user study.
\onecolumn

{\footnotesize
\begin{longtable}{p{0.05\textwidth}|p{0.2\textwidth}|p{0.08\textwidth}|p{0.06\textwidth}|p{0.3\textwidth}|p{0.08\textwidth}}
\caption{User Study Paper Analysis }
\label{metrics-user-study}\\
\hline
\multicolumn{6}{c}{\textbf{Hate Speech Moderation}} \\
\hline 
\textbf{Paper} & \textbf{Categories} & \textbf{Platforms} & \textbf{Size} & \textbf{Demographics} &\textbf{Study}  \\
\hline
\cite{schoenebeck2021youth}&Support, Removal Comprehension, Fairness and Bias&MyVoice&832&Only youth participants (14-24), 54.4\% females&Survey\\
\hline
\cite{schoenebeck2021drawing}&Effectiveness, Fairness and Bias&MTurk&573&Diverse demographics with transgender and non-binary participants, left-leaning (57\%)&Survey\\
\hline
\cite{jhaver2018online}&Effectiveness, Removal Comprehension, Fairness and Bias&Twitter&14&Mix of gender, mostly from US&Semi-structured interviews\\ \hline
\cite{blackwell2017classification}&Effectiveness&HeartMob&18&10 participants females, median age of participants was 40, high number of non-heterosexual participant&Semi-structured interviews\\ \hline
\cite{musgrave2022experiences}&Effectiveness, Fairness and Bias&Social media platforms and flyers&49&Black female participants, 30 participates from ages 18-30 years&Survey\\
\hline
\cite{thomas2022s}&Effectiveness&Residency program&135&56\% female, 99\% created content for YouTube, 30\% from the ages 35-44&Survey\\
\hline
\cite{sambasivan2019they}&Effectiveness&NGOs&199&Women participants from south-east Asia, 103 from India&Survey\\
\hline
\cite{doerfler2021m}&Effectiveness&Personal contacts&17&10 females, 4 LGBTQ+ participants, predominately white participants&Survey\\
\hline
\cite{gonccalves2021common}&Support, Removal Comprehension, Fairness and Bias&Dynata&2,870&Equal representations from US, Portugal and The Netherlands&Survey\\
\hline
\cite{xiao2022sensemaking}&Effectiveness, Support&College students&28&All participants from 18-20 years old, 18 participants were female, 23 students were of asian decent&Interviews\\
\hline
\multicolumn{6}{c}{\textbf{Misinformation Moderation}} \\ \hline
\cite{geeng2020fake}&Consumption of (fake/misinformation) news, Engagement of user&Facebook and Twitter& 25&One-third are college students, left leaning & Semi-structured interviews and Simulation \\ \hline
\cite{geeng2020social}&Consumption of (fake/misinformation) news, Support&Facebook and Twitter& 311 & Majority 18-24 year old (130),~38\% from UK, majority left leaning & Mixed-method survey\\ \hline
\cite{flintham2018falling}&Consumption of (fake/misinformation) news&Facebook&309 \color{red}(9)& Majority 18-25 year old (70\%), 55\% female, All from UK&Pilot study and simulation \\ \hline
\cite{kaiser2021adapting}&Effectiveness&MTurk&238 \color{red}(40)& Two-thirds male, over half of participants from 30-49 year old (92), left-leaning (159)&Pilot study and simulation\\ \hline
\cite{jahanbakhsh2021exploring}&Consumption of (fake/misinformation) news,Fairness and Bias&MTurk&1,807&47\% left-leaning, median age 35 year old, 42\% females&Pilot Study and survey\\ 
\hline
\cite{mena2020cleaning}&Engagement of users, Effectiveness&Facebook&501&51\% females, majority left-leaning (47.6\%), avg. age 36 year&Survey\\
\hline
\cite{clayton2020real}&Effectiveness&Facebook&2,994&Majority female (54\%), majority left-leaning (58\%)& Survey\\
\hline
\cite{sharevski2022misinformation}&Effectiveness&Twitter&319&56.4\% males,33.3\% from 25-34 age group, highly left-leaning (49.2\%)&Survey\\
\hline
\cite{gao2018label}&Effectiveness&MTurk&122&60\% females, almost equal number political leaning, ~80\% from 25 to 54 year old&Survey\\
\hline
\cite{saltz2020encounters}&Effectiveness, Support, Fairness and Bias&Dscout&23 \color{red}(15)&Equal genders,multiple ethnicities,balanced political views&Semi-structured interview and diary study\\
\hline
\cite{nyhan2020taking}&Effectiveness&Morning Consult and MTurk& 4,186 \color{red}(1,546)&54\% females, largely left-leaning (49\%), 42\% from 18 to 34 years old&Survey\\
\hline
\cite{seo2019trust}&Effectiveness&MTurk&800&55\% females, 75\% participants between the age of 20 to 40 years&Survey\\
\hline
\cite{epstein2021explanations}&Engagement of users, Effectiveness&Lucid&1,473& Mean age of participants were 47.87, 54.1\% were female&Survey\\
\hline
\cite{pennycook2020implied}&Engagement of users, Effectiveness&MTurk&5,271 \color{red}(1,568)&Female dominated (55\%), largely left leaning&Survey\\
\hline
\cite{ross2018fake}&Effectiveness&Facebook&151&57\% females, mean age was 25 year old&Survey\\
\hline
\cite{park2021experimental}&Engagement of users, Effectiveness&MTurk&11,145&52.9\% female participants&Pilot study and survey\\
\hline
\cite{sherman2021designing}&Consumption of (fake/misinformation) news&MTurk&1,456 \color{red}(24)&Almost equal number of male and females,large number of participants from 31-40 year of age&Pilot study and survey\\
\hline
\cite{urakami2022finding}&Consumption of (fake/misinformation) news&Twitter&15&66\% females, non-native English speakers&Survey\\
\hline
\cite{varanasi2022accost}&Consumption of (fake/misinformation) news&WhatsApp&28&53\% males, average age of 32&Survey\\
\hline
\cite{haque2020combating}&Consumption of (fake/misinformation) news&Facebook&519&74\% of the participants were from the 21-30 age range,64\% were students&Survey and interviews\\ \hline
\cite{lu2020government}&Consumption of (fake/misinformation) news, Effectiveness&WeChat&44&23 were male participants, 38 partipants were residing in China&Semi-structured interviews\\ \hline
\cite{jia2022understanding}&Engagement of users, Effectiveness&MTurk&1,677&Majority left-leaning 52\%, majority female 55\%&Simulation\\
\hline
\cite{grandhi2021crowd}&Effectiveness&MTurk&376&57\% females, 39\% from the ages 18-24 years, 52\% students&Survey\\
\hline
\cite{kirchner2020countering}&Effectiveness&Respondi&2,057&Mostly male participants&Survey, Semi-structured interview and simulation\\
\hline
\cite{seo2022if}&Engagement of users, Effectiveness&MTurk&2,841&Mostly female (52.3\%), largely between the ages of 28-37&Survey\\
\hline
\cite{yaqub2020effects}&Engagement of users, Effectiveness&MTurk&1,512&mostly male participants (51\%), slightly older crowd mean age 38&Survey\\
\hline
\multicolumn{6}{c}{\textbf{Moderation Techniques}} \\ \hline
\cite{riedl2021antecedents}&Support&U.S. national panel survey&1,022&53.2\% female, 75\% white ethnicity, 38.4\% from the ages 30-49&Survey\\
\hline
\cite{myers2018censored}&Support, Removal Comprehension, Fairness and Bias&OnlineCensors hip.org&519&Participants largely from US (295)&Survey\\
\hline
\cite{jhaver2019did}&Effectiveness, Support, Removal Comprehension, Fairness and Bias&Reddit&907&Participants from 81 countries, highest from US (61\%), majority male (81\%), under 25 year old (55\%)&Survey\\
\hline
\cite{haimson2021disproportionate}&Removal Comprehension, Fairness and Bias&Prolific, Qualtrics& 909 \color{red}(207)&Mixed genders, balanced ethnicity's, largely young, mix of conservatives and moderates& Survey\\
\hline
\cite{seering2019moderator}&Removal Comprehension& Twitch, Reddit, Facebook&56&Largly female and LGBTQ participants&Semi-structured interviews\\
\hline
\cite{eslami2019user}&Removal Comprehension&Yelp&15&40\% were between ages 35-44&Survey\\
\hline
\cite{bhuiyan2021nudgecred}&Consumption of (fake/misinformation) news&Twitter&430 \color{red}(12)&Slightly skewed towards females, higher number of independents and republicans&Pilot Study and Simulation\\
\hline
\cite{lu2021positive}&Consumption of (fake/misinformation) news&WeChat&33&More number of females (18), average age of 34&Semi-structured interviews\\
\hline
\cite{zhang2022shifting}&Consumption of (fake/misinformation) news&Qualtrics&177 \color{red}(21)&Largely female (66\%), between the ages of 25-34&Pilot study and survey\\
\hline
\cite{juneja2020through}&Removal Comprehension, Fairness and Bias&Reddit&13&70\% male population, 2 people from other countries other than US&Interviews\\
\hline
\cite{chung2022social}&Support&Prolific, Qualtrics, Embrain&5,392&Diverse population from US, UK, South Korea and Mexico, female dominated study (52.35\%), older participants (median age = 41)&Survey\\
\hline
\cite{duffy2022platform}&Support, Removal Comprehension, Fairness and Bias &TikTok, YouTube, Twitch&30&All participants had historically marginalized identities&Semi-structured interviews\\
\hline
\cite{lyons2022s}&Support, Fairness and Bias &MTurk&100&62\% male participants, all from USA, older participants (avg. age = 37)&Survey\\
\hline
\cite{suzor2019we}&Removal Comprehension, Fairness and Bias&OnlineCensors hip.org&380&N/A&Survey\\
\hline
\cite{blunt2020posting}&Support, Fairness and Bias &Twitter, Instagram&262&Females, 38.9\% were sex workers and activists&Survey\\
\hline
\end{longtable}
}
\begin{table*}
\centering
\caption{Papers about Detection Techniques}
\resizebox{\linewidth}{!}{%
\label{metrics-detection}
    \begin{tabular}{p{0.09\textwidth}|p{0.45\textwidth}|p{0.4\textwidth}}
    \hline
    \multicolumn{3}{c}{\textbf{Hate Speech Detection}} \\
    \hline
    \textbf{Categories}&\textbf{Broad Description}&\textbf{Papers} \\
    \hline
Content based& What type of features (i.e., TF-IDF, BoWV, POS tags etc.) can we obtain solely from the content to detect hate speech?  &~\cite{saha2018hateminers,davidson2017automated,salminen2018anatomy,de2018hate}\\ \hline
Lexicon Based& How can we utilize the syntactic and semantic orientations of the existing lexicons to identify hate speech? &~\cite{frenda2019online,vidgen2020detecting,capozzi2019computational,sanguinetti2018italian,basile2019semeval,tellez2018automated,vargas2021contextual,malmasi2017detecting,elsherief2018hate,wulczyn2017ex}\\ \hline
Deep Neural Based&How can we use the advancements in Deep Neural Networks to identify hate speech in platforms without using the handcrafted features ? &~\cite{yousaf2022deep,gamback2017using,alshamrani2021hate,chen2019use,zhang2018detecting,pitsilis2018detecting,agrawal2018deep,grondahl2018all,zhao2021comparative,malmasi2018challenges,doostmohammadi2020ghmerti,ribeiro2019inf,gomez2020exploring,winter2019know,kamble2018hate,rozental2019amobee,khan2020hateclassify,fortuna2020toxic,park2017one,badjatiya2019stereotypical,georgakopoulos2018convolutional,elnaggar2018stop,srivastava2018identifying,vaidya2020empirical,badjatiya2017deep,raut2022hate,ahmed2022deep,qian2018hierarchical,agarwal2021combating,ziems2005racism,li2022anti,vashistha2020online,modha2019ld,ousidhoum2019multilingual,mozafari2020hate,alatawi2021detecting,saha2019hatemonitors,caselli2020hatebert,devlin-etal-2019-bert,gpt_paper,liu2019roberta,wullach-etal-2021-fight-fire,founta2019unified,indurthi2019fermi,mozafari2019bert} \\ \hline
Hybrid Approaches&How can we effectively combine content based, lexicon based and deep neural based methods to identify hate speech?&~\cite{del2017hate,gao2017detecting,salminen2020developing,mathew2019thou,chatzakou2017mean,paschalides2020mandola,almerekhi2020these}\\
\hline
Multi-modal Based&How can we combine multiple modalities of posts (such as images, texts, memes) together to identify hate speech?&~\cite{yang2019exploring,singh2017toward,gomez2020exploring,kiela2020hateful,pramanick2021momenta,lippe2020multimodal,velioglu2020detecting,vijayaraghavan2021interpretable} \\
\hline
\multicolumn{3}{c}{\textbf{Misinformation Detection}} \\ \hline
Content Based & What type of features (i.e., TF-IDF, BoWV, POS tags etc.) can we obtain solely from the content to detect misinformation? &~\cite{singhal2021prevalence,zhou2020fake,horne2017just,della2018automatic,hosseinimotlagh2018unsupervised,potthast2018stylometric,rashkin2017truth,volkova2018misleading,wang2017liar,felber2021constraint,hakak2021ensemble} \\
\hline
Propagation Structure& How can we detect posts that contain misinformation by mining their spreading patterns in the underlying social networks?  &~\cite{wu2018tracing,sampson2016leveraging,shu2020hierarchical,liu2018early,mishra2020fake,singhal2021prevalence,shu2018understanding,shu2019role} \\ \hline
Hybrid Approaches&How can we effectively combine both content based features and propagation structure to detect misinformation? &~\cite{ruchansky2017csi,shu2019beyond,jin2017multimodal,karimi2018multi,helmstetter2018weakly}\\ \hline
Crowd Intelligence&Can we use the wisdom of the crowd to identify misinformation on the platform?&~\cite{kim2018leveraging,ma2018rumor,tschiatschek2018fake,vo2018rise,jin2016news,qian2018neural,shu2019defend}\\ \hline
Deep Neural Based&How can we use the advancements in Deep Neural Networks to identify misinformation in platforms without using the handcrafted features ?&~\cite{aloshban2020act,ren2020adversarial,bian2020rumor,yuan2021improving,ma2018rumor,shu2019defend,liu2020fned,ruchansky2017csi,monti2019fake,wang2018eann,kaliyar2021fakebert,song2021temporally,mehta2022tackling,xu2022evidence,cui2020deterrent,lu2020gcan,wu2018tracing,zimmerman2018improving} \\
\hline
Knowledge Based&Can we effectively use Knowledge Graphs to identify and automatically detect posts that contains misinformation?&~\cite{hu2021compare,kou2022hc,cui2020deterrent,dun2021kan,pan2018content,song2021temporally,mehta2022tackling,xu2022evidence}\\
\hline
Multi-modal Based&How can we combine multiple modalities of posts (such as images, texts, and audio) together to identify misinformation?&~\cite{singhal2022leveraging,khattar2019mvae,wang2018eann,wu2021multimodal,qi2019exploiting,shang2021multimodal,song2021multimodal}\\
\hline
\end{tabular}
}
\end{table*}

\begin{table*}
\centering
\caption{Data Driven Study Paper Analysis}
\resizebox{\linewidth}{!}{%
\label{metrics-data-study}
\begin{tabular}{p{0.04\textwidth}|p{0.13\textwidth}|p{0.09\textwidth}|p{0.12\textwidth}|p{0.10\textwidth}|p{0.09\textwidth}|p{0.12\textwidth}|p{0.12\textwidth}}
\hline
\multicolumn{8}{c}{\textbf{Hate Speech Moderation}} \\
\hline 
\textbf{Paper} & \textbf{Category} & \textbf{Platforms} & \textbf{Size of Dataset} & \textbf{Is Dataset Public?} &\textbf{Created the dataset?} & {\textbf{Type of Data used}} & {\textbf{Intervention Studied}} \\
\hline
\cite{chandrasekharan2020quarantined}&Engagement of users, Effectiveness&Reddit&$\sim$85M&No&Yes&Posts and metadata&Soft Moderation\\
\hline
\cite{ribeiro2020does}&Engagement of users, Effectiveness&Reddit&$\sim$6.3M&Yes&Partially&Comments, posts and metadata&Hard Moderation\\ \hline
\cite{chandrasekharan2017you}&Engagement of users, Effectiveness&Reddit&$\sim$100M&Partially&No&Comments and posts&Hard Moderation\\ \hline
\cite{ali2021understanding}&Engagement of users, Effectiveness&Gab, Reddit and Twitter&$\sim$30M&Partially&Partially&Posts and metadata&Hard Moderation\\ \hline
\cite{jhaver2021evaluating}&Engagement of users, Effectiveness&Twitter&$\sim$49M&No&Yes&Tweets and metadata&Hard Moderation\\ \hline
\cite{saleem2018aftermath}&Engagement of users, Effectiveness&Reddit&$\sim$1.9M&Partially&No&Comments and metadata&Hard Moderation\\ \hline
\cite{chandrasekharan2018internet}&Effectiveness&Reddit&$\sim$2.8M&No&Yes&Comments&Hard Moderation\\ \hline
\cite{munger2017tweetment}&Effectiveness&Twitter&N/A&No&Yes&Tweets and metadata&Hard Moderation\\ \hline
\cite{trujillo2022make}&Engagement of users, Effectiveness&Reddit&$\sim$15M&Yes&No&Posts&Soft Moderation\\ \hline
\color{red}\cite{rauchfleisch2021deplatforming}&Engagement of users, Effectiveness&YouTube and BitChute&$\sim$11K&No&Partially&Videos and metadata&Hard Moderation\\ \hline
\cite{rogers2020deplatforming}&Effectiveness&Telegram&N/A&No&Yes&Posts and metadata&Hard Moderation\\ \hline
\color{red}\cite{kumarswamy2022strict}&Effectiveness&Parler&$\sim$200M&Partially&Partially&Posts and metadata&Hard Moderation\\ \hline
\cite{shen2022tale}&Engagement of users, Effectiveness&Reddit&$\sim$3.7M&Partially&No&Posts, metadata&Soft Moderation\\

\hline
\multicolumn{8}{c}{\textbf{Misinformation Moderation}} \\ \hline
\cite{zannettou2021won}&Engagement of users&Twitter&$\sim$18K&Yes&Yes&Tweets and metadata&Soft Moderation\\
\hline
\cite{ling2022learn}&Engagement of users&TikTok&$\sim$41K&No&Yes&Videos and metadata&Soft Moderation\\
\hline
\cite{kim2020effects}&Engagement of users, Effectiveness&YouTube&105&No&Yes&Videos metadata&Soft Moderation\\
\hline
\cite{papakyriakopoulos2022impact}&Engagement of users, Effectiveness&Twitter&$\sim$2.4M&Yes&Yes&Tweets and metadata&Soft Moderation\\
\hline
\multicolumn{8}{c}{\textbf{Moderation Techniques}} \\ \hline
\cite{shen2019discourse}&Fairness and Bias&Reddit&$\sim$9K&Yes&No&Posts&Soft Moderation\\ \hline
\cite{thomas2021behavior}&Effectiveness&Reddit&N/A&Partially&No&Posts and comments&Hard Moderation\\
\hline
\color{blue}\cite{juneja2020through}&Removal Comprehension, Fairness and Bias&Reddit&$\sim$0.5M&No&Yes&Moderation logs&Hard Moderation\\
\hline
\end{tabular}
}
\end{table*}

\end{document}